\documentclass{elsart}

 \usepackage{graphics}
 \usepackage{graphicx}
 \usepackage{epsfig}

\usepackage{amssymb, amsmath}

\begin{document}

\begin{frontmatter}




\title{High-energy cosmic ray fluxes in the Earth atmosphere: calculations vs experiments}


\author{A. A. Kochanov, T. S. Sinegovskaya, and S. I. Sinegovsky}

\address{Irkutsk State University, Irkutsk, Russia}

\begin{abstract}

A new calculation of the atmospheric fluxes of cosmic-ray hadrons and muons in the energy
range $10$--$10^5$ GeV has been performed for the set of  hadron production models, 
EPOS 1.6, QGSJET II-03, SIBYLL 2.1, and others that are of interest to cosmic ray physicists.  
The fluxes of secondary cosmic rays at several levels in the atmosphere are computed using directly data of the ATIC-2, GAMMA experiments, and the model proposed recently by Zatsepin and Sokolskaya as well as the parameterization of the primary cosmic ray spectrum by Gaisser and Honda. The calculated energy spectra of the hadrons and muon flux as a function of zenith angle are compared with measurements as well as other calculations.  
The effect of uncertainties both in the primary cosmic ray flux and  hadronic model predictions on the spectra of atmospheric hadrons and muons is considered. 

\end{abstract}

\end{frontmatter}

\section{Introduction}

A comparison of the calculated spectra and zenith-angle distributions of atmospheric hadrons and muons with the results of measurements makes it possible to solve related problems: (i) indirect research of the high-energy behavior of the  primary cosmic ray spectrum and composition provided that cross sections of hadron-nuclear interactions are known, and 
(ii) a study of the high-energy hadron-nucleus interactions provided that energy spectra and composition of primary cosmic rays or spectra of secondaries are measured with a satisfactory accuracy.

In this work we analyze the possibility to discriminate indirectly hadronic interaction models leaning on recent measurements of hadron and muon fluxes. We present new calculations of atmospheric cosmic-ray fluxes in the range $10$--$10^5$ GeV performed with current models extensively used in the simulation of high-energy cosmic ray propagation through the atmosphere, QGSJET01~\cite{QGSJET01}, QGSJET II~\cite{QGSJET2, ostap08}, SIBYLL~\cite{SIBYLL}, NEXUS~\cite{NEXUS}, EPOS~\cite{EPOS} and  Kimel and Mokhov (KM)~\cite{KMN}. 
Numerical results are obtained using directly data for the primary cosmic ray spectra and composition of the ATIC-2 ~\cite{ATIC2} and GAMMA~\cite{GAMMA} experiments. Alternatively we use recently proposed model by Zatsepin and Sokolskaya (ZS)~\cite{ZS3CM} as well as the parameterization by Gaisser and Honda  (GH)~\cite{GH}.  The comparison of the results of these calculations with recent direct measurements of high-energy atmospheric muons~\cite{CAPRICE94, BESSTEV, L3C, Lecoultre} and  with predictions of other authors allow to evaluate uncertainties caused by statistical errors of the primary spectrum data as well as to discriminate them from those originated from hadronic models.

The calculations are based on the method originally developed for solving problems of neutrino transport in matter~\cite{Naumov99} and  modified subsequently~\cite{NS} to apply for the transport problem of the nucleon component of cosmic rays in the Earth's atmosphere. The solution of the nucleon-meson cascade equations was stated in~\cite {KSS, KPSS}. The method allow to solve numerically nuclear-cascade equations for an arbitrary primary spectrum and for the most general form of hadron production cross sections. 
Here we do not focus upon detailed comparison with the results of other calculations, referring readers to the works~\cite{ZK60,VZK79, Dedenko89, Lipari, Bugaev98, Naumov2001, VZ01} and reviews~\cite{Ryazhskaya96, Costa, Naumov2004, Hebbeker02} where one can find the relevant experimental data and comparison of calculations performed by other authors and a comprehensive list of references. 
  
The plan of this work is as follows.
In sections 2 and 3 we touch briefly the primary cosmic-ray spectra and modern high-energy hadronic models. In section 4 we describe the method to solve the problem of meson transport through the atmosphere. In section 5 we calculate the hadron fluxes and compare them  to old and new measurements having particular interest to test the calculation scheme. In sections 6 and 7 we discuss the impact of hadronic models and the primary spectra on the sea-level muon flux at different zenith angles. In the short summary, we touch upon the uncertainties of current muon flux predictions resulted from the hadronic models and primary spectra. 

\section{Primary cosmic ray spectra}
   
In our calculations we put primary emphasis on recent data on the primary cosmic ray (PCR) spectra and composition obtained with Advanced Thin Ionization Calorimeter, balloon-borne experiment ATIC-2~\cite{ATIC2}. In order to compare the predictions with the high-energy measurements of the AM flux we extend the calculations to higher energies, up to $100$ TeV, using also the PCR spectrum data of the GAMMA experiment~\cite{GAMMA}.  The PCR model by Zatsepin and Sokolskaya~\cite{ZS3CM}, supported by the ATIC-2 data, was applied as the reliable instrument to extrapolate median-energy data to high-energy ones.  Besides we use the parameterization of PCR spectra by Gaisser and Honda~\cite{GH}.  
\begin{figure}[b!]
\begin{center}
\includegraphics[width=0.95\textwidth, trim = 0.5cm 0.5cm 0.0cm 0.2cm]{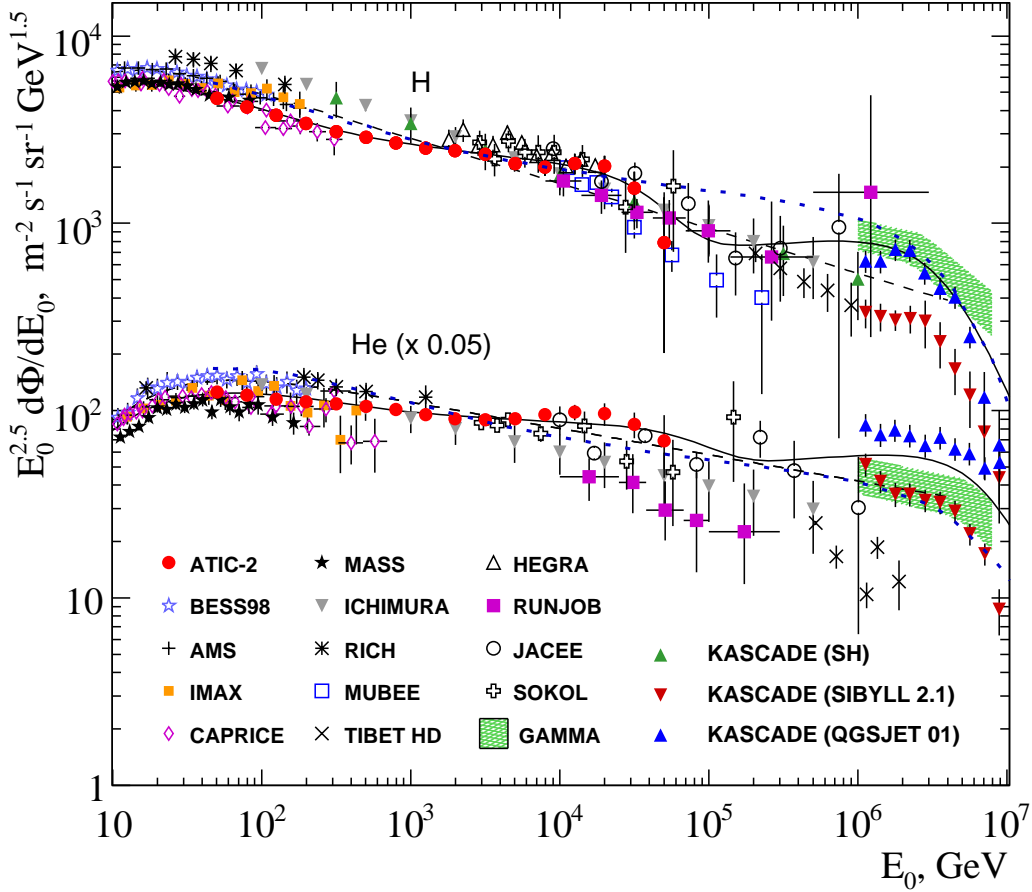}
\end{center}
\caption{Primary proton and helium spectra combining
  balloon, satellite and ground-based measurements. The solid curves present the ZS model~\cite{ZS3CM}, dashed -- the GH parameterization~\cite{GH} and dotted line -- BV calculations~\cite{Berezhko}.}  
		\label{PCR}
\end{figure}

The ATIC-2 experiment designed to measure primary cosmic ray spectra with the individual charge resolution from proton to iron in the wide energy interval $50$ GeV -- $200$ TeV enabled to obtain the data with a high statistical assurance.  Proton and helium spectra measured in the ATIC-2 experiment have different slopes and differ from a simple power law (Fig.~\ref{PCR}). In this figure shown are also the energy spectra of protons and helium nuclei obtained in the balloon, satellite and ground based experiments~\cite{BESS98, AMS, IMAX, CAP98, MASS, RICH, MUBEE, RUNJOB, JACEE, SOKOL,  KASSH, KAS_elemnt, HEGRA, TIBETHD, ICHIMURA}. 
The ATIC-2 data agree with the data of magnetic spectrometers (BESS~\cite{BESS98}, AMS~\cite{AMS}, IMAX~\cite{IMAX}, CAPRICE~\cite{CAP98}, MASS~\cite{MASS}) up to $300$ GeV. In the energy region $1<E<10$ TeV, the ATIC-2 data are consistent with the SOKOL~\cite{SOKOL} measurements and those of atmospheric Cherenkov light detector HEGRA~\cite{HEGRA}. At energies above $\sim 10$ TeV, the spectra become steeper, and follow the data of emulsion chamber experiments MUBEE~\cite{MUBEE} and JACEE~\cite{JACEE}, though the agreement is not so clear.

The KASCADE experiment data for proton spectrum~\cite{KASSH} in the energy range $ 3\times10^2$--$10^6$ GeV are deduced from the single-hadron (SH) spectrum assuming contributions from helium and heavy nuclei  subtracted (green 
upward triangles in Fig.~\ref{PCR}).  The later KASCADE measurements~\cite{KAS_elemnt} for elemental groups of cosmic rays via the detection of extensive air showers (EAS) at energies above $10^6$ GeV  (near and above the knee in the primary spectrum) are presented here by the primary proton and helium energy spectra (extracted directly from Figs. 14, 15 in Ref.~\cite{KAS_elemnt}). These spectra were obtained on the base of EAS simulations with two hadronic interaction models, QGSJET01 (solid upward triangles) and SIBYLL 2.1 (solid downward triangles). 


Solid curves in Fig.~\ref{PCR} present the model suggested by Zatsepin and Sokolskaya (ZS)~\cite{ZS3CM} that fits well the ATIC-2 experimental data and describes PCR spectra in the energy range $10$--$10^7$ GeV. The dashed lines show the GH parameterization~\cite{GH}. In order to extend our calculation to higher energies, the PCR spectra measured in the GAMMA experiment was used. The energy spectra and elemental composition, obtained in the GAMMA experiment, cover the $10^3$--$10^5$ TeV range (shaded bands) and agree with the corresponding extrapolations of known balloon and satellite data at $E\geq10^3$ TeV. In the present calculation employed was the version of the GAMMA spectra reconstructed in the framework of $1,2$D combined analysis with the SIBYLL 2.1 model (see~\cite{GAMMA} for details).

The ZS model comprises contributions to the cosmic ray flux of three classes of astrophysical sources like supernova and nova shocks. In Fig.~\ref{PCR} also plotted are  the PCR spectra (dotted lines) calculated by Berezhko \& V$\ddot{\text o}$lk (BV)~\cite{Berezhko} with use of nonlinear kinetic theory for diffusive acceleration of  charged particles in supernova remnants in which the magnetic field is amplified by the efficiently accelerating nuclear CR component.  The BV model gives a natural explanation for the observed knee in the Galactic CR spectrum.

 
The BV spectra and ZS ones are close in the energy interval $10$--$10^{5}$ GeV  inspite of the difference in shape, reflecting basic assumptions of the models. The difference in proton spectra becomes more appreciable at the energies $10^{5}$-- $10^{6}$ while the difference in helium spectra is revealed above $10^{6}$ GeV.
Therefore one may expect that both ZS primary spectra and BV ones would lead to similar muon fluxes in wide energy range. 

At very high energies the BV spectra agree rather well with the KASCADE protons (QGSJET 01) and helium (SIBYLL 2.1) as well as protons and helium spectra derived from GAMMA experiment with usage of SIBYLL 2.1. 
 Note also that some discrepancy exists between the KASCADE proton spectum and GAMMA one, both obtained with usage SIBYLL 2.1. 
\begin{figure}[b!]
	\centering
\vspace{0.42cm}
\includegraphics[width=0.43\textwidth, trim = 1.1cm 0.0cm 0.0cm 0.0cm]{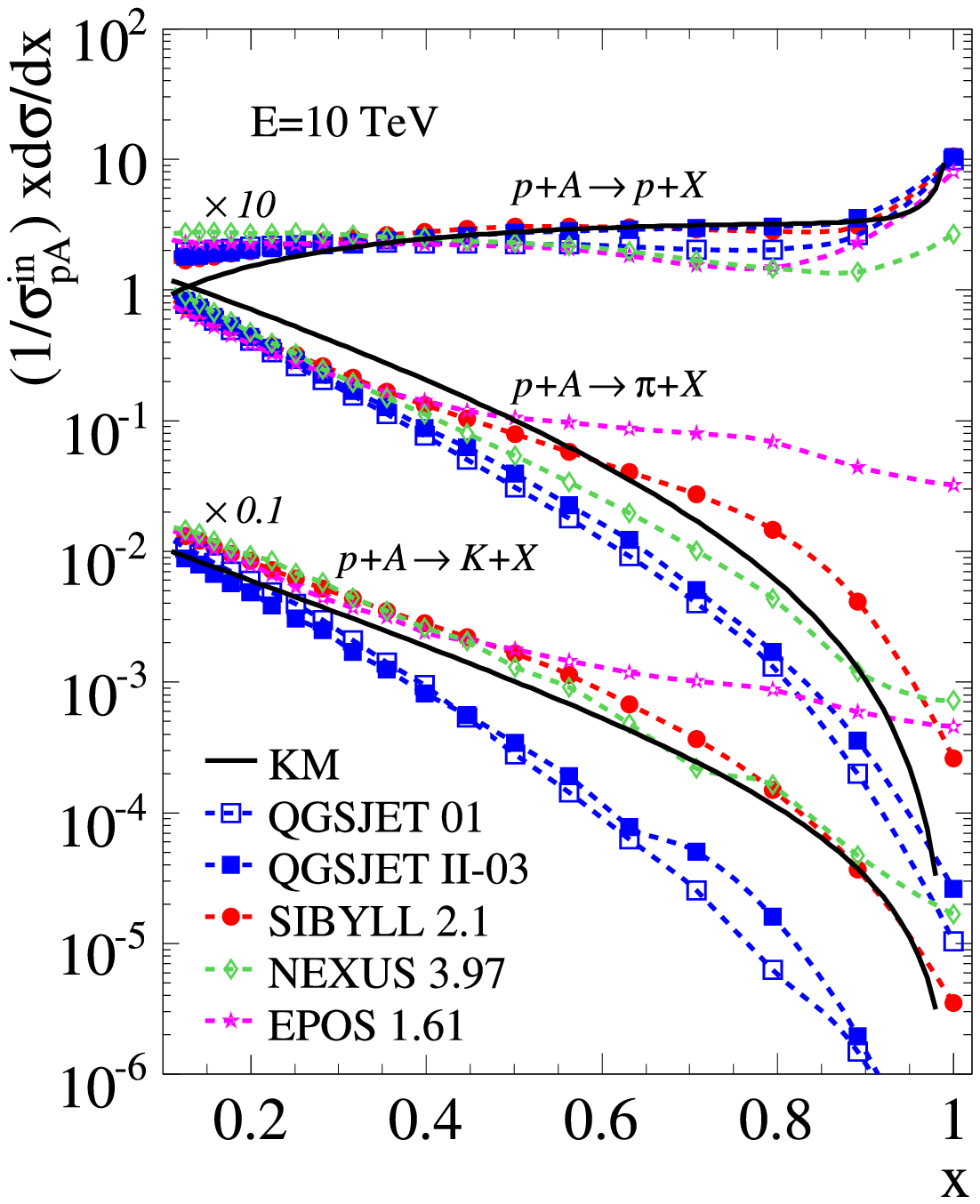}
\hspace{1.2 cm}
\includegraphics[width=0.43\textwidth, trim = 1.1cm 0.0cm 0.0cm 0.0cm]{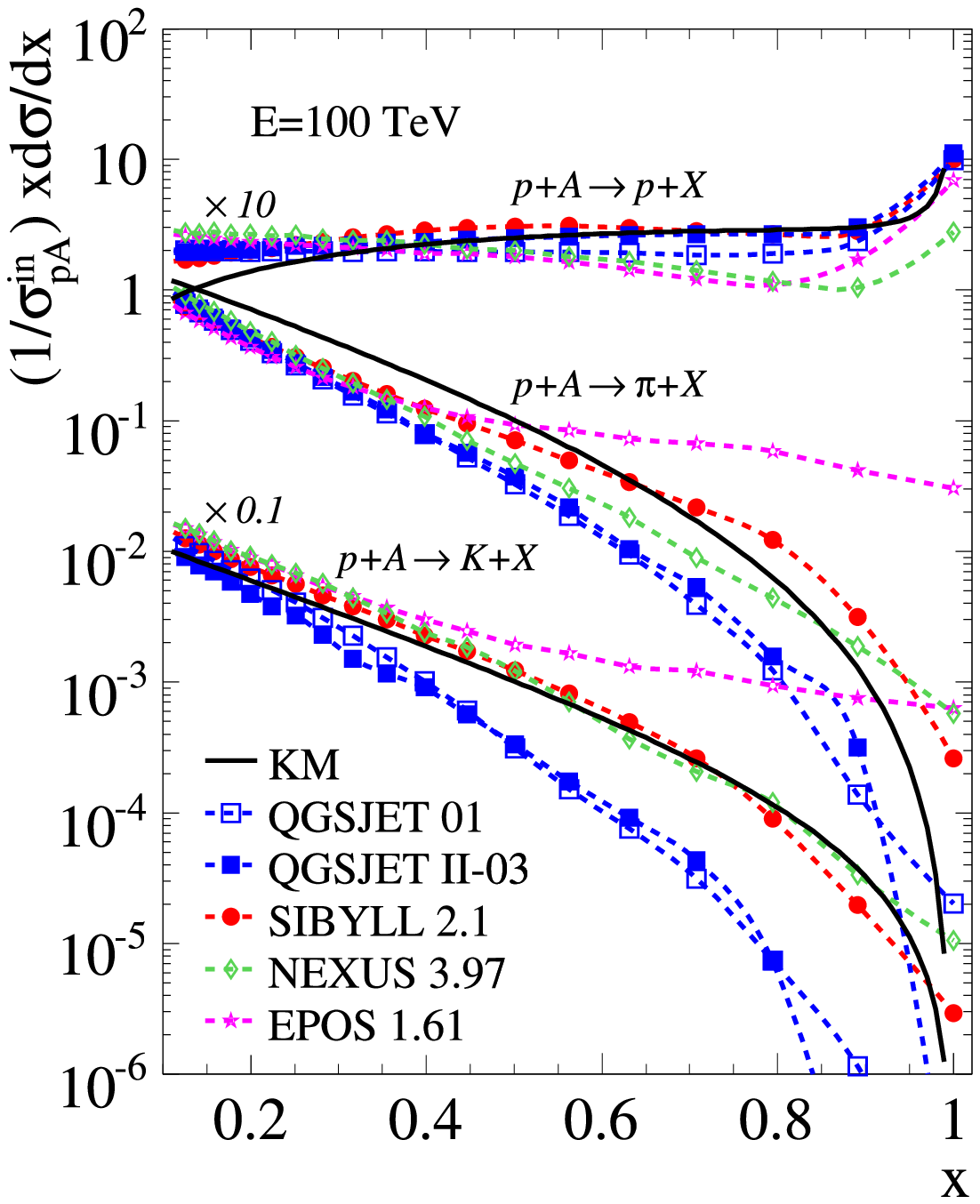}
\caption{Spectra of the particle production in proton-air collisions at $E_0=10$ TeV (left) and $100$ TeV (right) according to the currently employed hadronic interaction models, KM, QGSJET, SIBYLL, NEXUS, and EPOS.}  
	\label{had_models}
\end{figure}

\section{Hadronic interaction models}

\begin{figure}[t!]
	\centering
\includegraphics[width=0.90\textwidth, trim = 0.6cm 0.0cm 0.5cm 0.0cm]{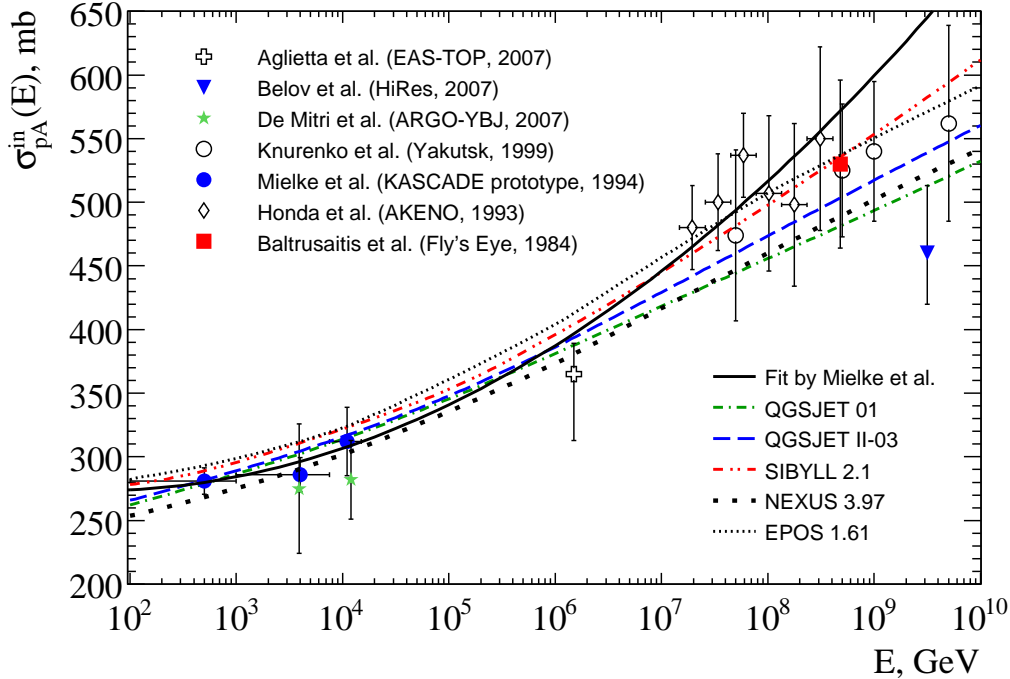}
\caption{Proton-air inelastic cross-section: the model predictions vs. experiments~\cite{Mielke, Baltrusaitis, Honda, Yakutsk, HiRes, EASTOP1, ARGO}. Solid line presents the fit $\sigma_{pA}^{\rm in}=290-8.7\log E+1.14\log^2 E$   ~\cite{Mielke}. 
}  
 \label{pAinel}
\end{figure}

To present day the direct measurements of inclusive cross sections for the nucleon and meson production in hadron-nucleus collisions have been extended up to energies of about 1 TeV only.
Therefore cosmic-ray calculations require either developing a well-justified procedure for extrapolating cross sections measured at moderate energies or developing the models that would give reliable predictions at high and ultrahigh energies. 

In our studies, we apply several hadronic interaction models, QGSJET 01~\cite{QGSJET01},  QGSJET II-03~\cite{QGSJET2}, SIBYLL 2.1~\cite{SIBYLL}, NEXUS 3.97~\cite{NEXUS}, and EPOS~\cite{EPOS}, that was successfully tested in the modern simulation programs CORSIKA~\cite{CORSIKA}, AIRES~\cite{aires} and CONEX~\cite{CONEX} to analyze data on extensive air showers. Another hadronic model we use is the original one proposed by Kimel and Mokhov~\cite{KMN}, for which we take updated parameters~\cite{NS, Naumov2001, Naumov2004}. 
Based on accelerator data, obtained at energies up to $\sim 1$ TeV, this model is compatible both in shape and magnitude with the current high energy hadronic models (Fig.~\ref{had_models}). The KM model was also applied in 3D Monte Carlo calculations of the atmospheric muon and neutrino fluxes~\cite{Derome}.

Prominent features of $h$A-interactions are the violation of Feynman scaling at high energy and the growth of total inelastic cross sections with energy (Fig.~\ref{pAinel}). Figure~\ref{pAinel} presents the hadron model predictions for  the proton-air inelastic cross section along with experiments~\cite{Mielke, Baltrusaitis, Honda, Yakutsk, HiRes, EASTOP1, ARGO}. Solid line here shows the fit to the data obtained with the prototype of the KASCADE hadron calorimeter~\cite{Mielke}.

 As illustrated in Fig.~\ref{had_models}, the current models differ in the predicted amount of the scaling violation in the energy range  $10$--$100$ TeV. For example, in the central region ($x < 0.1$) QGSJET II predicts the growth of the inclusive cross section for reactions $pA \rightarrow pX$ 
by a factor of 1.4 and for EPOS scaling violation is about 1.5. Apparently, this do not significantly affect the numerical results, because the contribution of small $x$ to the fluxes of secondary hadrons is negligible due to falling energy spectrum of primary cosmic rays. Unlike the central region, the scaling violation in the fragmentation region (say, at $x \sim 0.5-0.9$) is most important, in particular in case of EPOS model: this one predicts a decrease of inclusive cross sections for  $pA \rightarrow pX$ up to $15\%$,  while the EPOS pion and kaon yield is sizeably amplified (see Fig.~\ref{had_models} and Table \ref{tab_z}).  

To illustrate the difference of above hadron models the spectrum-weighted moments was computed for proton-air interactions $pA\rightarrow cX$ with the inclusive spectra
 $F_{pc}(E_0, E_c)\equiv(x/\sigma^{in}_{pA})\,{d\sigma_{pc}}/{dx}$:
\begin{equation}\label{mom}
z_{pc}(E_0)
= \int\limits_0^1x^{\gamma-1}F_{pc}(E_0, xE_0)\,dx
=\int\limits_0^1\frac {x^{\gamma}}{\sigma^{in}_{pA}}\frac {d\sigma_{pc}}{dx}\,dx,
\end{equation}
where  $x=E_c/E_0$, \ $c=p,n,\pi^\pm, K^\pm$. As one can see from the Table \ref{tab_z}, $z_{pc}$ demonstrate approximate scaling law  of  SIBYLL 2.1 and KM models, while in the case of QGSJET II and EPOS a sizeable violation of the scaling is found, particularly for the latter (see also~\cite{ostap08, Lipari}). 
\begin{table*}[t]
\protect\caption[Z-factors]
{Spectrum-weighted moments $z_{pc}$ calculated for $\gamma=1.7$. \label{tab_z}}
		\begin{tabular}{l c c c c c}\hline
Model  & $E_0$, TeV & $z_{pp}$ & $z_{pn}$ & $z_{p\pi^+}+z_{p\pi^-}$ & $z_{pK^{\tiny +}}+z_{pK^-}$\\\hline
		QGSJET II  & 0.1 & 0.174 & 0.088 & 0.075 & 0.0062\\
						   & 1   & 0.198 & 0.094 & 0.064 & 0.0061\\
						   & 10  & 0.205 & 0.090 & 0.059 & 0.0061\\\hline
				 EPOS  & 0.1 & 0.199 & 0.112 & 0.058 & 0.0067\\
				 		   & 1   & 0.167 & 0.084 & 0.083 & 0.0099\\
						   & 10  & 0.147 & 0.073 & 0.083 & 0.0139\\\hline
		SIBYLL 2.1 & 0.1 & 0.211 & 0.059 & 0.068 & 0.0148\\
							 & 1   & 0.209 & 0.045 & 0.070 & 0.0143\\
							 & 10  & 0.203 & 0.043 & 0.068 & 0.0124\\\hline
		           &    &$z_{pp}$&$z_{pn}$  & $z_{p\pi^+}  \qquad z_{p\pi^-}$
		                 & $z_{pK^+} \qquad z_{pK^-}$\\
		KM         & 0.1 & 0.178 & 0.060 & 0.044 \quad 0.0027 & 0.0051  \quad  0.0015     \\
							 & 1   & 0.190 & 0.060 & 0.046 \quad 0.0028 & 0.0052  \quad  0.0015   \\
							 & 10  & 0.182 & 0.052 & 0.046 \quad 0.0029 & 0.0052  \quad  0.0015  \\\hline					 
				 
\end{tabular}
\end{table*}
%

\section{Transport of cosmic-ray mesons}

Here we concentrate on the meson component of the atmospheric hadron cascade, since the nucleon component was studied in detail~\cite{NS}. At sufficiently high energies, there is no need to consider 3D effects of the cascade, the electromagnetic energy loss of hadrons, and the effect of the geomagnetic field. The collisions of cosmic-ray nuclei are treated with the semisuperposition model (SSP)~\cite{EGLS92} (see also~\cite{Lipari08}) according to which the interaction of a nucleus ($A, Z$) with energy $E_A$ is similar to the averaged superposition of $Z$ proton and $A-Z$ neutrons interactions each with energy $E=E_A/A$.
The SSP model predicts the same average values of additive observable quantities as the superposition model but the fluctuations.   



%
%

The meson component of the cascade decouples from the nucleon component if one neglects the small contribution of $N\overline{N}$ pair production processes in meson-nucleus interactions.
Under these assumptions, a set of integro-differential equations for charged-pion transport can be written as
\begin{eqnarray} \label{skpi}\nonumber
\frac{\partial \pi^{\pm} (E,h,\vartheta)}{\partial h} &=&
-\frac{\pi^{\pm}(E,h,\vartheta)}{\lambda_{\pi} (E)}
-\frac{m_\pi \pi^{\pm}(E,h,\vartheta)}{p {\tau}_\pi \rho (h,\vartheta)}\\
&+& \sum _i G_{i \pi^{\pm}}^{\mathrm{int}} (E,h,\vartheta)+
\sum _K G_{K \pi^{\pm}}^{\mathrm{dec}} (E,h,\vartheta)+ \\ \nonumber
&+&\frac{1}{\lambda_{\pi}(E)}\int\limits_{E}^{\infty}
\frac{1}{ \sigma_{\pi A}^{\mathrm{in}}(E) }
\frac{d\sigma_{\pi^{\pm}\pi^{\pm}}(E_0, E)}{dE}\pi^{\pm}(E_0,h,\vartheta)dE_0+	\\ \nonumber  
&+&\frac{1}{\lambda_{\pi}(E)}\int\limits_{E_{\pi}^{\mathrm{min}}}^{\infty}
\frac{1}{ \sigma_{\pi A}^{\mathrm{in}}(E) }
\frac{d\sigma_{\pi^{\mp}\pi^{\pm}}(E_0,E)}{dE}\pi^{\mp}(E_0,h,\vartheta)dE_0,
\end{eqnarray}
where $\pi^{\pm}(E,h,\theta)$  is the flux of charged pions with energy $E$ at the slant depth $h$ (in g\,cm$^{-2}$) in the atmosphere along a direction of zenith angle $\theta $, $\sigma_{\pi A}^{\mathrm{in}}(E)$ is the inelastic-interaction cross section, $\lambda_{\pi}(E)=[N_0\sigma_{\pi A}^{\mathrm{in}}(E)]^{-1}$ is the pion interaction length ($N_0=N_A/A$), $p$,  $m_{\pi}$, and $\tau_{\pi}$  are momentum, mass and lifetime of the pion, $d\sigma(E_0, E)/dE$ is the cross section of charged pion production in $h$A-collisions, $\rho(h,\theta)$ is the air density profile, parameterized according to the Linsley's model of the US standard atmosphere (see~\cite{aires} for details).
The initial conditions for Eqs.~(\ref{skpi}) are $\pi^{\pm}(E,h=0,\theta)=0 $.

Sources of pions are strong interactions $i+A=\pi^{\pm}+X\ (i=p,\ n,\ K^{\pm},\ K^0,\ \bar K^0)$ and the weak decays of kaons, $K=K^{\pm},\ K^0_L$, and $K^0_S$. Hence two types of the pion source functions are 
\begin{equation}
\label{GNPI}
G_{i \pi^{\pm}}^{\mathrm{int}} (E,h,\vartheta) =\frac{1}{\lambda_i(E)}
                                   \int\limits^{\infty }_{E_{i}^{\min}}
                                   \frac{1}{\sigma _{iA}^{{\mathrm{in}}}(E)}
                                   \frac{d\sigma _{i\pi^\pm}(E_0, E)}
                                   {dE} D_i(E_0,h,\vartheta) dE_0,
\end{equation}
\begin{eqnarray}  \label{GKPI} 
G_{K \pi^{\pm}}^{\mathrm{dec}} (E,h,\vartheta) & &= B(K_{2\pi})
                               \frac{m_K}{\tau _K \rho (h,\vartheta)}
                          		 \int\limits_{E^{\rm min}_{K_{2\pi}}}^{E^{\rm max}_{K_{2\pi}}}
                               \frac{dE_0}{p_0^2} F_{K_{2\pi}}^\pi (E_0, E)
                                 K(E_0,h,\vartheta)              \\ \nonumber
						                    & &  + B(K^0_{\ell 3})
                              \frac{m_K}{\tau _K \rho (h,\vartheta)}
                              \int\limits_{E^{\rm min}_{K_{\ell3}}}^{E^{\rm max}_{K_{\ell3}}}                             	\frac{dE_0}{p_0^2}F_{K^0_{\ell3}}^\pi(E_0, E)
                                 K^0(E_0,h,\vartheta).                               
\end{eqnarray}
Here $B(K_{2\pi})$ and $B(K^0_{\ell 3})$ are the branching ratios for the decays $ K^\pm \rightarrow \pi^\pm \pi^0$, $ K^0_S \rightarrow \pi^+ \pi^- $, and $ K^0_{\ell3} \rightarrow \pi^\pm \ell ^\mp \nu_\ell$. $E_0$, $p_0$,  $m_K$, and $\tau_K$ are the kaon energy, momentum, mass, and lifetime. $~F_{K_{2\pi}}^\pi$ and $F_{K^0_{\ell3}}^\pi$ are the $\pi$-meson production spectra in the two- and three-particle kaon decays (see~\cite{NC98, NCC98}).
The function $D_i(E_0,h,\vartheta)$ in the right-hand side of (\ref{GNPI}) and the function $K(E_0,h,\vartheta)$ in (\ref{GKPI}) are the fluxes of the parent particles $i$ and $K$. The differential nucleon fluxes $D_p(E_0,h)$ and $D_n(E_0,h)$ were calculated using the formulas from the work~\cite{NS}. The processes of the pion regeneration and inelastic charge exchange are accounted by the last two terms in Eqs.~(\ref{skpi}). The integration limits in Eqs.~(\ref{skpi}), (\ref{GNPI}), and (\ref{GKPI}) are given in~\cite{KSS}.

At the first step we neglect a small contribution of kaons to the pion flux. This will be done at the second step after calculation of the kaon component.
Let's denote by $\tilde{\pi}^\pm $ the pion flux in the absence of the kaon source and introduce the combinations  
\begin{equation}
 \Pi^\pm(E,h,\vartheta)=\tilde{\pi}^+(E,h,\vartheta){\pm} \ \tilde{\pi}^-(E,h,\vartheta),     
\end{equation}
which obey the equations 
\begin{eqnarray} \label{skupi}
\frac{\partial \Pi ^{\pm} (E,h,\vartheta)}{\partial h} &=&
-\frac{\Pi ^{\pm}(E,h,\vartheta)}{\lambda_{\pi} (E)}
-\frac{m_\pi \Pi ^{\pm}(E,h,\vartheta)}{p {\tau}_\pi \rho (h,\vartheta)}
+G^{\pm}_{N\pi} (E,h,\vartheta)+ \nonumber \\
&&+ \frac{1}{\lambda_{\pi} (E)} \int\limits^{1}_0
\Phi ^{\pm}_{\pi \pi} (E,x) \Pi ^{\pm} (E/x,h,\vartheta) \frac{dx}{x^2}\ , 
\end{eqnarray}
where $ x=E/E_0$  and
\begin{multline}\nonumber
G^{\pm}_{N\pi} (E,h,\vartheta)=\left[G_{p \pi^+}^{\mathrm{int}} (E,h,\vartheta)+G_{n \pi^+}^{\mathrm{int}}(E,h,\vartheta) \right] \pm \\
\left[G_{p \pi^-}^{\mathrm{int}} (E,h,\vartheta)+G_{n \pi^-}^{\mathrm{int}} (E,h,\vartheta)\right],
\end{multline}
$$
\Phi_{\pi\pi}^\pm (E,x) = \frac{E}{\sigma_{\pi A}^{{\mathrm{in}}}(E)}
\left[\frac{d\sigma_{\pi^+\pi^+}(E_0, E)}{dE} \pm
\frac{d\sigma_{\pi^+\pi^-}(E_0, E)}{dE}\right]_{E_0=E/x} .
$$
Following~\cite{NS} we introduce the basic function of the method, $\cal Z$-factor,  which depends on the variables $E, h,\vartheta$  
\begin{equation}
	 {\cal Z}_{\pi \pi}^{\pm }(E,h,\vartheta)=
\int\limits^1_{0} \Phi _{\pi \pi}^\pm (E,x)\frac{ \Pi ^{\pm} (E/x,h,\vartheta)}
{\Pi ^{\pm} (E,h,\vartheta)}\frac{dx}{x^2}\ ,
\end{equation}
 and recast Eq.~(\ref{skupi}):
\begin{eqnarray} \label{piZ}
\frac{\partial \Pi ^{\pm} (E,h,\vartheta)}{\partial h} &=&
-\frac{\Pi ^{\pm}(E,h,\vartheta)}{\lambda_{\pi} (E)}
-\frac{m_\pi \Pi ^{\pm}(E,h,\vartheta)}{p {\tau}_\pi \rho (h,\vartheta)}
+G^{\pm}_{N\pi} (E,h,\vartheta)+ \label{zpi} \nonumber \\
&&+ \frac{1}{\lambda_{\pi} (E)} {\cal Z}_{\pi \pi}^{\pm }(E,h,\vartheta)
\Pi ^{\pm} (E,h,\vartheta) \ .
\end{eqnarray}
 The function ${\cal Z}(E, h,\vartheta)$ serves as a generalization of the  
CR spectrum-weighted moments $z_{if}(\gamma)$ (see e.g.~\cite{Lipari, Bugaev98, VNS}) to the case of non-power cosmic ray spectrum, violation of Feynman scaling, and energy-dependent inelastic cross sections.

Formally a solution of Eq.(\ref{piZ}) is:
\begin{equation}
\begin{split}
	\Pi ^{\pm}(E,h,\vartheta)&= \int\limits^{h}_{0} dt~G^{\pm}_{N\pi}(E,t,\vartheta)\\
								&\times\exp \left [- \int\limits^{h}_{t} dz 
                    \left (\frac{1-{\cal Z}_{\pi \pi}^ \pm (E,z,
                    \vartheta)}{\lambda_\pi (E)}+ \frac{m_\pi }
                   {p {\tau}_\pi \rho (z,\vartheta)} \right ) \right ].
\end{split}                 
\end{equation} 
Then we solve via iterations the equations with unknown $\cal Z$-factors.  In zero-order approximation, we neglect the pion regeneration and charge exchange,  
 $ {\cal Z} _{\pi \pi}^{\pm (0)}(E,h)= 0$. %
The differential energy spectrum of pions calculated under this assumption 
\begin{equation}
\label{P0}
\Pi ^{\pm (0)}(E,h,\vartheta) = \int\limits^{h}_{0} dt~G^{\pm}_{N \pi} (E,t,\vartheta)\exp\left [- \int\limits^{h}_{t} dz
  \left ( \frac{1}{\lambda_\pi (E)}+ \frac{m_\pi }{p {\tau}_\pi
\rho (z,\vartheta)} \right ) \right ]	
\end{equation}
enables one to obtain $\cal Z$-factors in the first-order approximation. In the first-order and next approximations, the regeneration and charge exchange processes are already involved:
\begin{equation}\label{Pn}
\begin{split}
\Pi ^{\pm (n)}(E,h,\vartheta) &= \int\limits^{h}_{0} dt
                          ~G^{\pm}_{N \pi}(E,t,\vartheta) \\ 
                        &\times\exp \left [- \int\limits^{h}_{t} dz
                        \left ( \frac{1-{\cal Z}_{\pi \pi}^{\pm (n)}
                        (E,z,\vartheta)}{\lambda_\pi (E)}+
                        \frac{m_\pi }{p {\tau}_\pi \rho (z,\vartheta)}
                        \right ) \right ].
\end{split}                      	
\end{equation}
In the ($n$+$1$)-iteration, the value of ${\cal Z}$ is calculated using the pion flux, obtained at the $n$-th step:
\begin{eqnarray}
{\cal Z}_{\pi \pi}^{\pm (n+1)}(E,h,\vartheta) = \int\limits^{1}_{0} \Phi _{\pi \pi}^{\pm }
(E,x) \frac{\Pi ^{\pm (n)}(E/x,h,\vartheta)}
{x^2 \Pi ^{\pm (n)}(E,h,\vartheta)} dx. 
\end{eqnarray}

Now the process of $N\overline{N}$ pair production in pion-nucleus collisions 
can be taken into account as a correction to the nucleon flux. In turn,  the pion flux is refined taking into consideration this additional source.

The evolution equations for charged and neutral kaons in the atmosphere  have the form 
\begin{eqnarray} \label{skr}
\frac{\partial K (E,h,\vartheta)}{\partial h} &=&
-\frac{K(E,h,\vartheta)}{\lambda_K (E)}
-\frac{m_K K(E,h,\vartheta)}{p {\tau}_K \rho (h,\vartheta)}
+\sum _{i=p,n,\pi} G_{i K} (E,h,\vartheta) + \nonumber      \\
& & + \frac{1}{\lambda_K (E)}
\int\limits^{\infty }_{E^{\min} _{KK}}
\frac{1}{\sigma _{K A}^{{\mathrm{in}}}(E)} \frac{d\sigma _{KK} (E_0, E)}
{dE} K(E_0,h,\vartheta) dE_0,
\end{eqnarray}
where $G_{i K} (E,h,\vartheta)$ are the source functions of kaons produced in $N$A- and
 $\pi$A-interactions: 
\begin{eqnarray}
G_{i K} (E,h,\vartheta)& =& \frac{1}{\lambda_i(E)} \int\limits^{\infty }_{E_{NK}^{\min}}
                  \frac{1}{\sigma_{iA}^{{\mathrm{in}}}(E)}
                  \frac{d\sigma _{iK}(E_0, E)}{dE}D_i(E_{0},h,\vartheta)dE_0\ . 
\end{eqnarray}
These equations are similar to Eqs.(\ref{skupi}), therefore we solve them using the same procedure. Finally we have
\begin{eqnarray}
\label{Kn}
K^{(n)}(E,h,\vartheta) &=& \int\limits^{h}_{0} dt~\sum_{N, \pi}G_{iK}(E,t,\vartheta)\times \nonumber \\            
                       & &\times \exp \left [- \int\limits^{h}_{t} dz
                        \left ( \frac{1-{\cal Z}_{KK}^{(n)}
                        (E,z,\vartheta)}{\lambda_K (E)}+
                        \frac{m_K}{p {\tau}_K \rho (z,\vartheta)}
                        \right ) \right ].	
\end{eqnarray}

At last, we use the solutions, obtained for the kaon spectra, to take into account the  contribution to the pion flux of the kaon source. 
 
\section{Hadron fluxes}

\begin{figure}
	\centering
\includegraphics[width=0.95\textwidth, trim = 0.5cm 0.0cm 0.3cm 0.0cm]{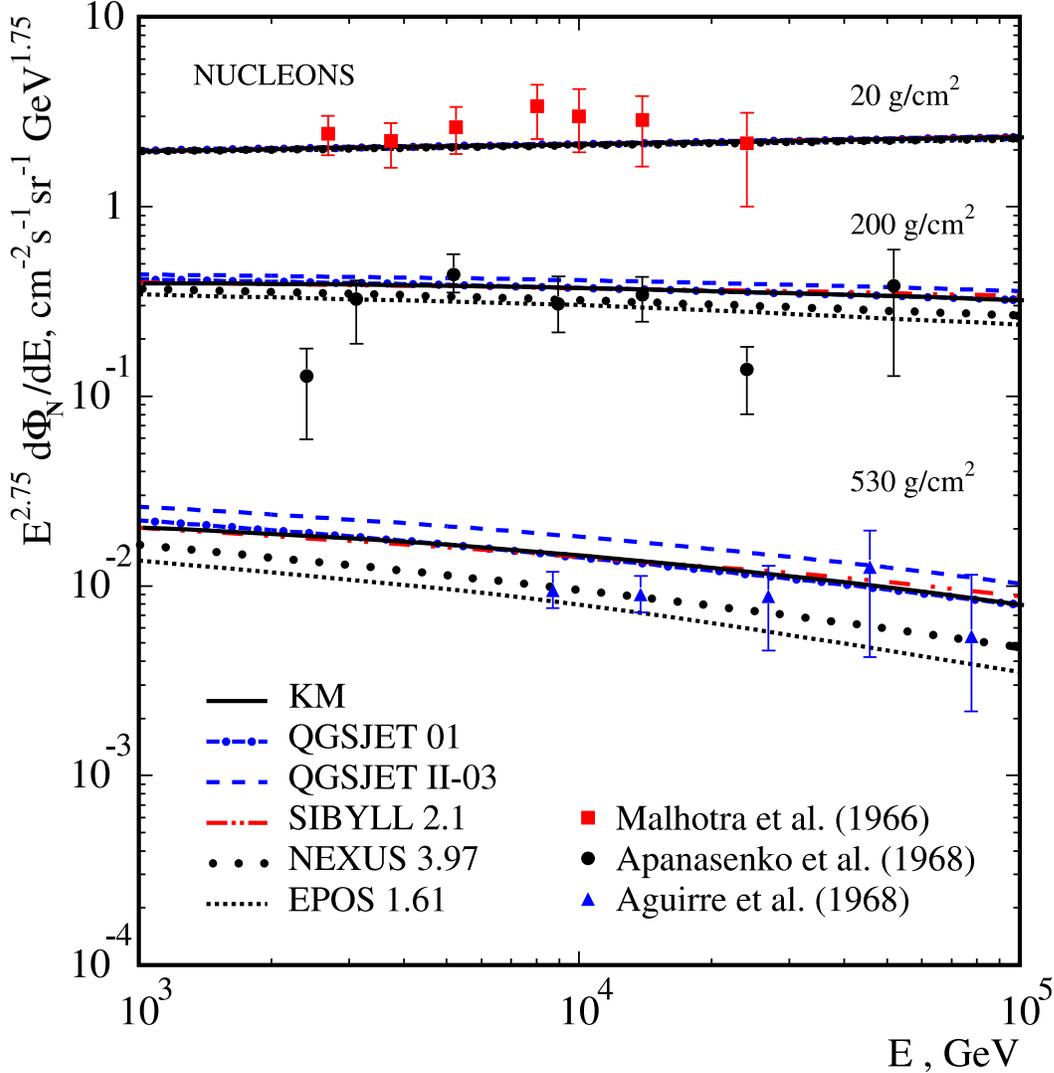}
	\caption{Nucleon energy spectra at three atmospheric depths, computed with the GH primary spectrum and various hadronic models. The experimental data are from Refs.~\cite{Malhotra, Apanasenko, Aguirre}.
}  
  		\label{NUCLEONS}
\end{figure}

The solution of the nucleon and meson transport equations discussed above enables us to compute the hadron fluxes with hadronic models under consideration and compare them to experiments.
Fig.~\ref{NUCLEONS} shows the differential intensity of nucleons in the atmosphere calculated   for the GH primary spectrum along with the measurement data~\cite{Malhotra, Apanasenko, Aguirre} at depths of $20$, $200$, and $530$ g\,cm$^{-2}$.  Evidently at small atmospheric depths there is no strong dependence of the flux on the hadronic interaction model but the difference accumulates as depth increases. 
The data shown in Fig.~\ref{NUCLEONS} being obtained
with rather large  experimental errors do not afford a possibility to distinguish between this work predictions. The calculations performed with QGSJET II result in the maximum  nucleon flux whereas  EPOS leads  the minimum one. The rest of hadronic models spread between them.

\begin{figure}
	\centering
\includegraphics[width=0.95\textwidth, trim = 0.5cm 0.0cm 0.3cm 0.0cm]{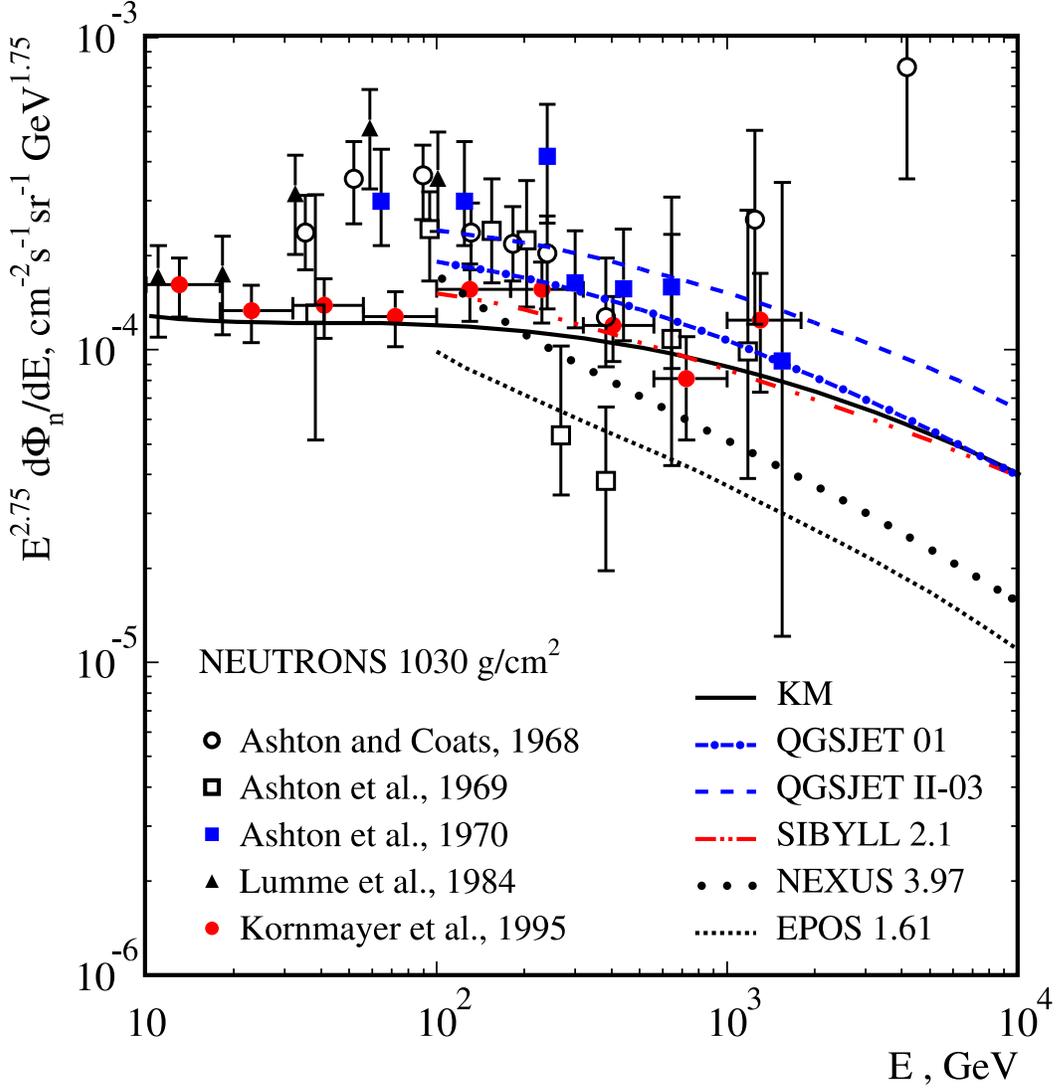}
	\caption{Neutron energy spectra at sea level. The curves present this work calculations with usage of diverse hadronic models and the GH primary cosmic-ray spectrum. The experimental data are taken from Ref.~\cite{KASCADE}.
}  
  		\label{NEUTRONS}
\end{figure}

Experimental data on the neutron component at the sea level are more detailed but rather contradictory. In Fig.~\ref{NEUTRONS}, we compare the calculated energy spectra of neutrons at the sea level with the data taken from ref.~\cite{KASCADE}. 
The calculations with QGSJET01, KM and SIBYLL models result in best agreement with most recent data obtained from a hadron calorimeter prototype of the KASCADE facility~\cite{KASCADE}. One may hope that further experiments to study the high-energy nucleon flux will allow a more detailed test of hadronic models.

\begin{figure}
	\centering
\includegraphics[width=0.95\textwidth, trim = 0.4cm 0.0cm 0.3cm 0.0cm]{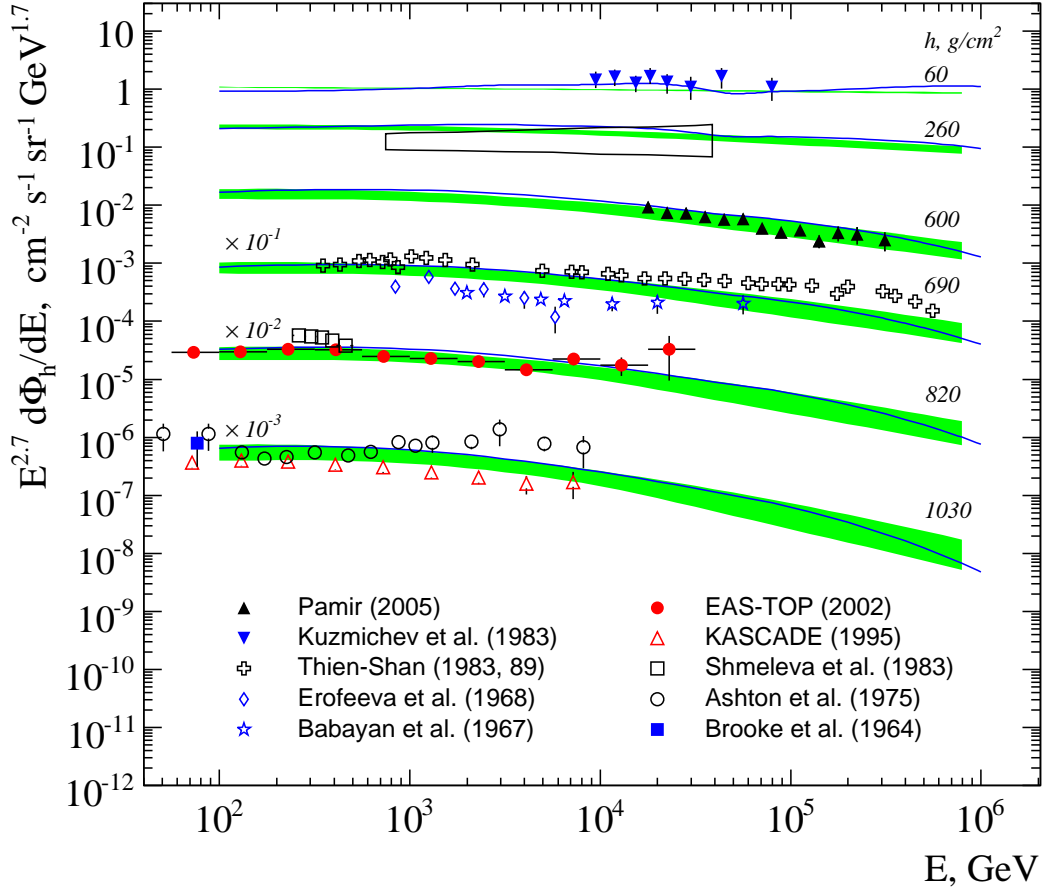}
\caption{All hadron energy spectra in the atmosphere. Solid lines present this work calculation with the QGSJET II hadronic model and ATIC-2 primary cosmic-ray spectrum. The bands show the spread of our predictions due to the usage of diverse hadronic models for the GH primary spectrum. The experimental data are from Refs.~\cite{KASCADE, Kuzmichev, Takahashi, Malinowski, ThienShan, Erofeeva, Babayan, EASTOP, Shmeleva, Ashton}.
}  
 \label{ALLHADRONS}
\end{figure}

Fig.~\ref{ALLHADRONS} shows the calculated spectra of hadrons compared with the measurements \cite{KASCADE, Kuzmichev, Takahashi, Malinowski, ThienShan,  Erofeeva, Babayan, EASTOP, Shmeleva, Ashton}. The hadron fluxes are computed at several depths of the atmosphere in the range  $60$--$1030$ g\,cm$^{-2}$  using  the  QGSJET II hadronic model, the ATIC-2 data, and    the ZS cosmic ray spectrum in the region of higher energies (solid lines). Predictions for the GH parameterization combined with the QGSJET01, QGSJET II, SIBYLLL, NEXUS,  and  EPOS models  are shown as shaded areas. Upper bounds of the bands correspond to the predictions of QGSJET II and lower ones -- to those of EPOS, whereas the other models give the hadron intensities  between the above  predictions. Calculation results are scaled with the factors shown at the left side of the figure. 

At shallow depths there are only few experimental data sets available. For a depth of $60$ g\,cm$^{-2}$, we compare our results with the data of the Moscow group~\cite{Kuzmichev} obtained onboard balloons with X-ray emulsion chambers.  For a depth of $260$ g\,cm$^{-2}$  we use the data~\cite{Takahashi} (shown as rectangle area) obtained on the airplane with help of emulsion chamber also. In both the cases, our results well agree with the data within the experimental errors.  

\begin{figure}
	\centering
\includegraphics[width=0.95\textwidth, trim = 0.5cm 0.0cm 0.4cm 0.0cm]{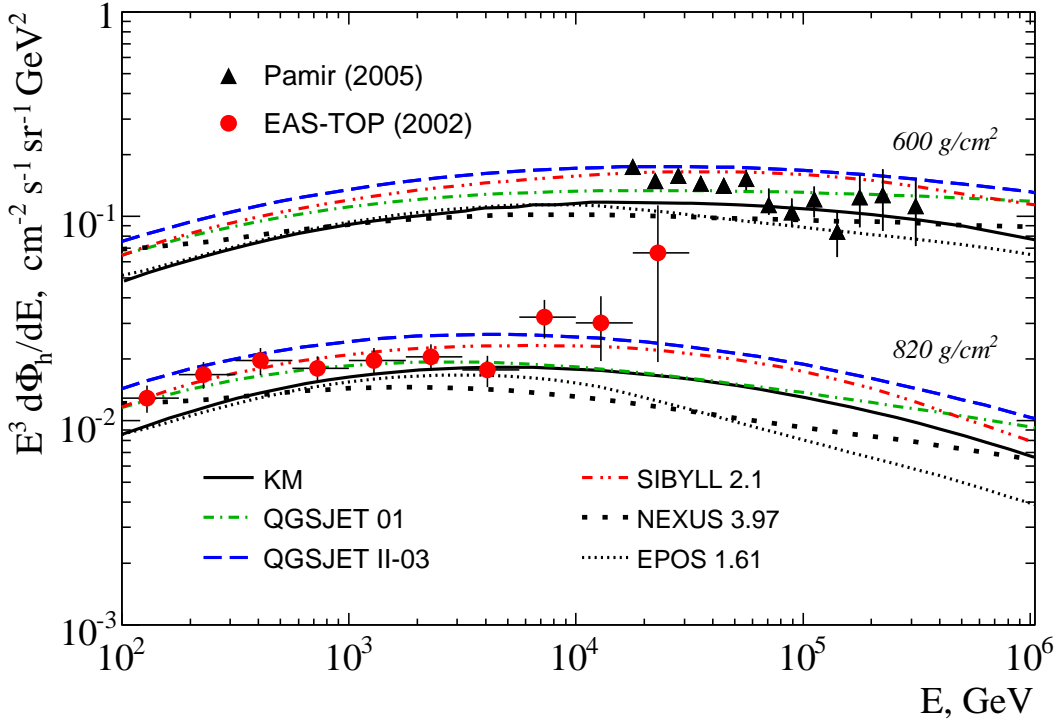}
	\caption{Scaled data of the Pamir and EAS-TOP experiments. Lines  present the calculations  for the GH primary spectra using every hadronic model under study. 
}  
  		\label{PAMIR}
\end{figure}

For atmospheric depths of $600$--$700$ g\,cm$^{-2}$, we compare our predictions with  measurements performed at the Thien-Shan and Pamir. High-energy data obtained with the carbon emulsion chambers in the Pamir experiment~\cite{Malinowski} agree well with our calculations, while the data from the ionization calorimeter of the Thien-Shan experiment~\cite{ThienShan} give the hadron intensity at $E>50 $ TeV about twice as high as our prediction with QGSJET II model.
For the atmospheric depth $820$ g\,cm$^{-2}$, our calculations agree rather well with the EASTOP data~\cite{EASTOP}. For $1030$ g\,cm$^{-2}$ our calculations at energies above TeV are in some discrepancy with the data of the KASCADE experiment~\cite{KASCADE} as well as with those of early measurements (taken from~\cite{KASCADE}). 
Also in Fig.~\ref{PAMIR} we compare in detail our predictions for each of considered hadronic models with the recent data of the Pamir and EAS-TOP experiments.

\begin{figure}
	\centering
\includegraphics[width=0.9\textwidth, trim = 0.0cm 0.0cm 0cm 0.0cm]{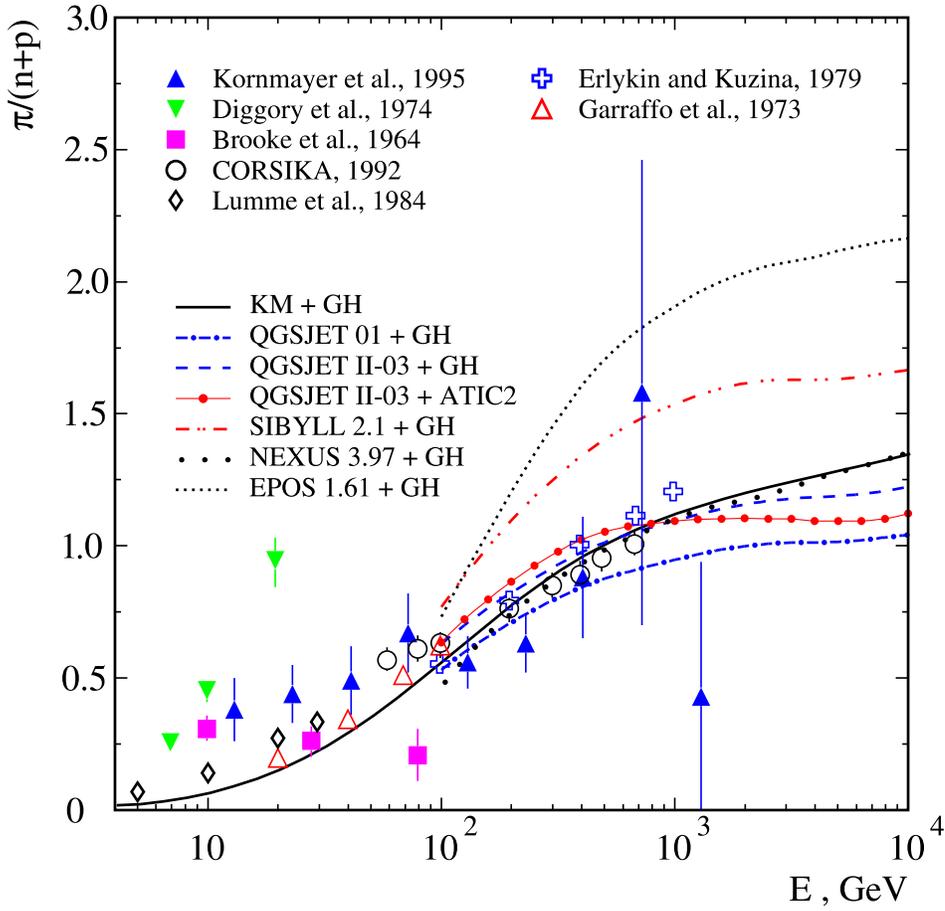}
	\caption{The ratio $\pi$/(p+n) calculated at sea level with the set of hadronic models. 
The ATIC-2 primary spectrum and GH parameterization are used.	Experimental data (closed symbols) and calculations of other works (open symbols) are taken from Ref.~\cite{KASCADE}.
}  
  		\label{pinratio}
\end{figure}

\begin{figure}
	\centering
	\hspace{0.5cm}
\includegraphics[width=0.47\textwidth, trim = 0.9cm 0.0cm 0.0cm 0.0cm]{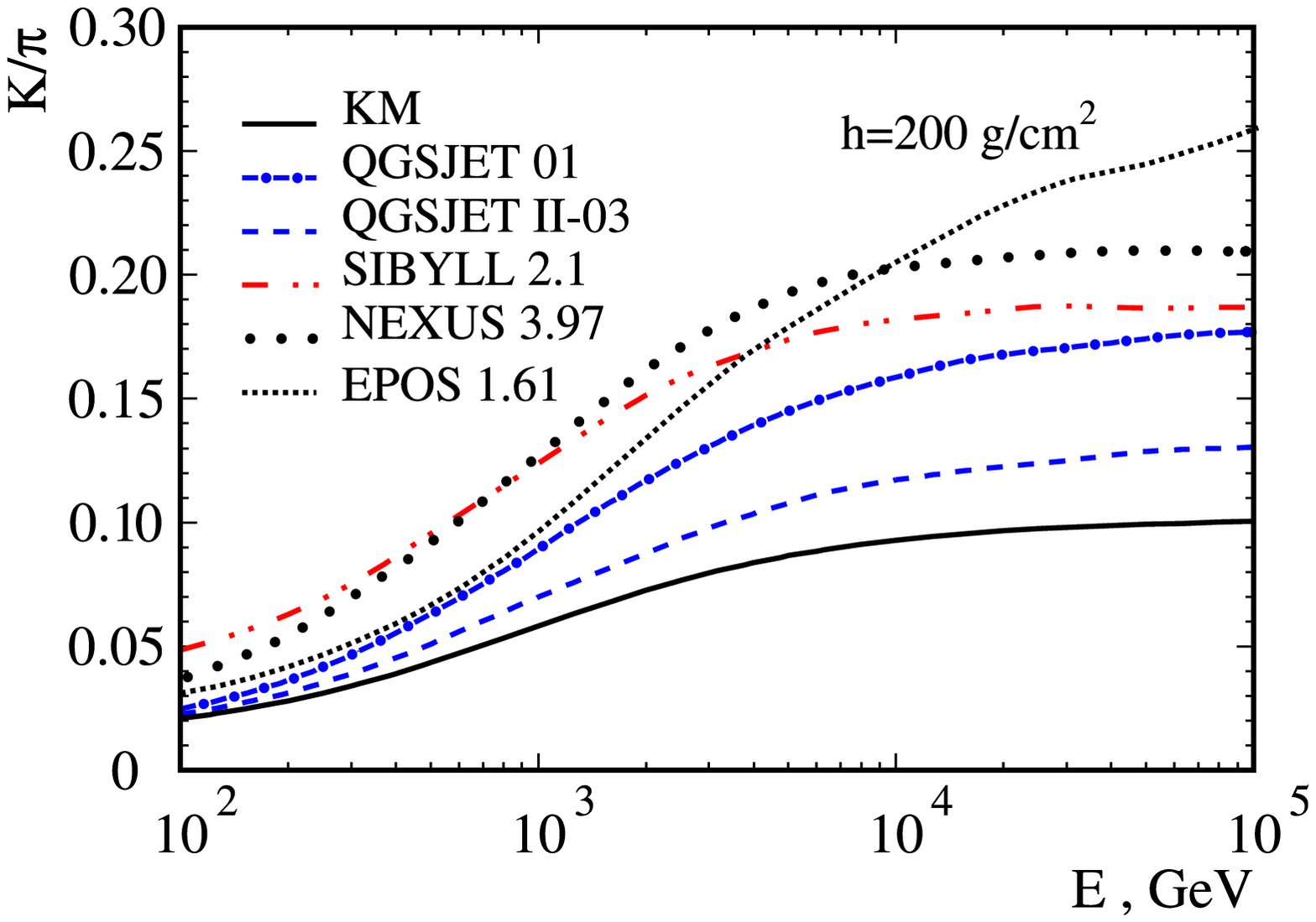}
\hspace{0.10cm}
\includegraphics[width=0.46\textwidth, trim = 0.9cm 0.0cm 0.0cm 0.0cm]{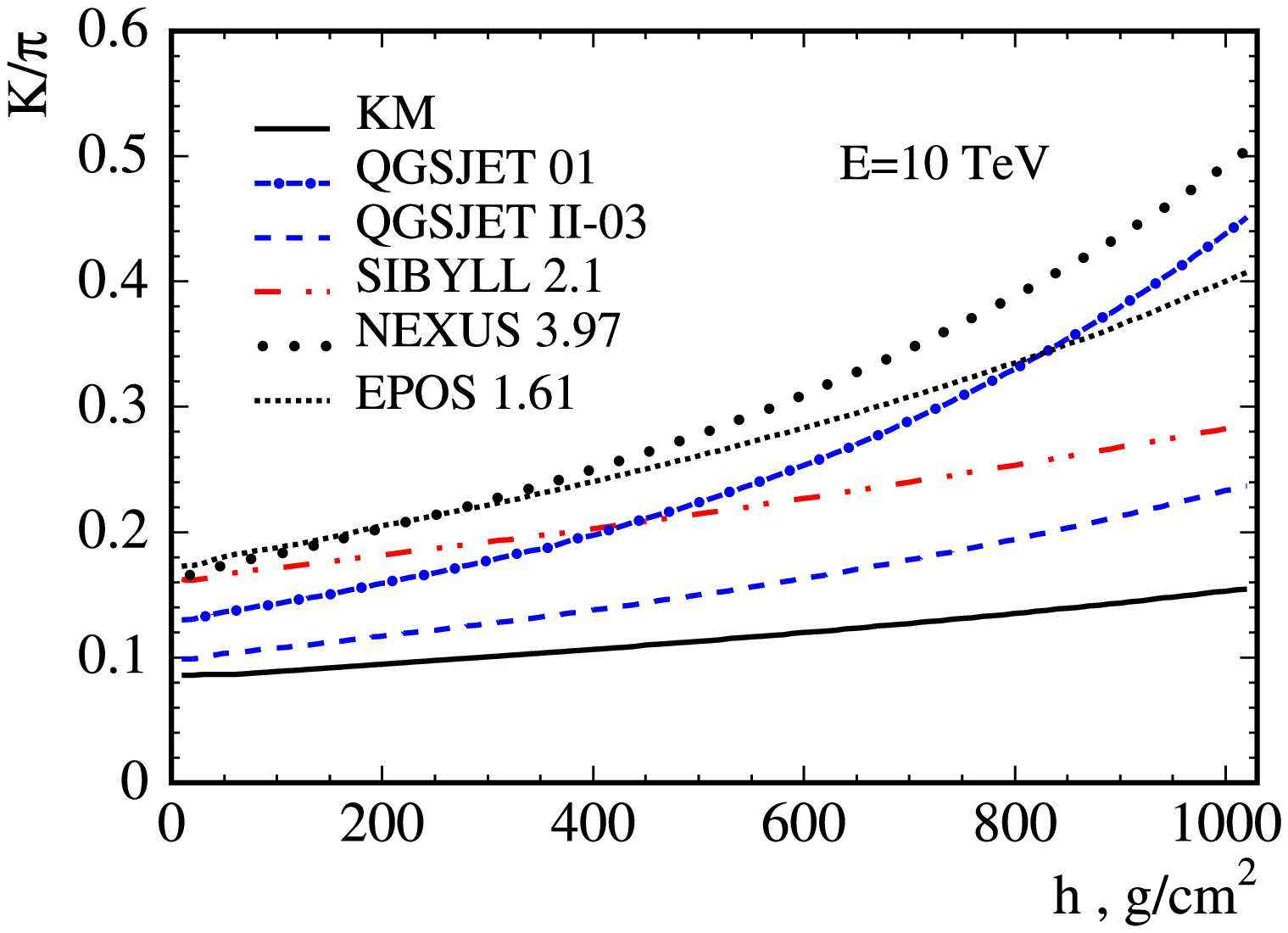}
	\caption{Charged $K/\pi$-ratio, depending on the energy,  calculated for the atmospheric depth of 200 g\,cm$^{-2}$ (left plot), and that at $10$ TeV unfolded along the atmospheric depth (right plot). GH primary spectrum is used.
}  
  		\label{kpiratio}
\end{figure}

In Fig.~\ref{pinratio} we plot the calculated ratio of the pion flux to nucleon one  ($\pi/N$ ratio) at sea level together with experimental results and other calculations taken from~\cite{KASCADE}. As one can see, the primary spectra and mass composition weakly affect the ratio unlike  hadronic models. This allows in some way to test the cross sections: the models KM, NEXUS, QGSJET 01, and QGSJET II yield comparatively similar results, whereas the models SIBYLL and EPOS  lead to the sharp difference in $\pi/N$ ratio.  
The left panel of Fig.~\ref{kpiratio} shows the charged $K/\pi$ ratio of the fluxes calculated for the atmospheric depth $200$ g\,cm$^{-2}$. It is seen that the NEXUS and EPOS  predict maximal $K/\pi$ ratios $\sim 0.2$--$0.25$ at $E=100$ TeV, whereas the KM and QGSJET II give minimal values, $\sim 0.1$--$0.15$. The right panel of Fig.~\ref{kpiratio} shows the same ratio unfolded along the atmospheric depth.

\section{Muon fluxes} 

The muons are abundantly generated in decays of the mesons that are produced through hadron-nuclei collisions in the atmosphere. Apart from the dominant source of atmospheric muons, $\pi_{\mu2}$ and  $K_{\mu2}$ decays, we take into consideration three-particle semileptonic decays, $K^{\pm}_{\mu3}$, $K^{0}_{\mu3}$, and small contributions originated  from decay chains  $K\rightarrow\pi\rightarrow\mu$ ~($K^0_S\rightarrow \pi^+ + \pi^-$, $K^\pm \rightarrow \pi^\pm +\pi ^0$, $K^0_{L}\rightarrow \pi^\pm+ \ell^\mp+ \bar{\nu}_{\ell}(\nu_{\ell})$, $\ell=e, \mu $).
At very high energies, the bulk of muons is expected to arise from decays of the charmed particles, $D$ and $\Lambda_c$. 

\begin{figure}[b!]
	\begin{center}
\includegraphics[width=0.95\textwidth, trim = 2.5cm 0.5cm 1.8cm 0cm]{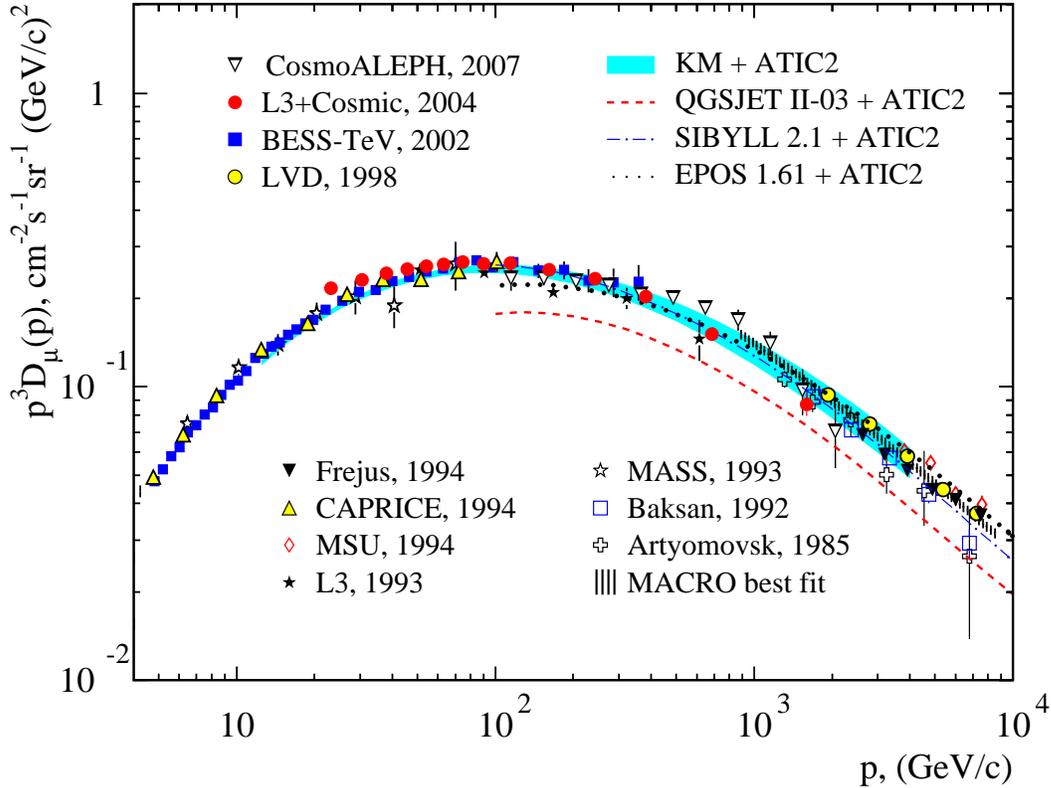} %
	\end{center}
\caption{Energy spectrum of muons at ground level near the vertical. 
The curves show this work calculations for the  ATIC-2 primary cosmic-ray spectrum and different hadronic models. The band presents a result for the KM model taking into consideration statistical errors of the ATIC-2 data. The experimental data are from~\cite{CAPRICE94, BESSTEV, L3C, Lecoultre,  Bugaev98, MASS93, MSU, MACRO, LVD, Frejus, Baksan, ASD}.}
	\label{mu-1}
\end{figure}

\begin{figure}
	\begin{center}
\includegraphics[width=0.95\textwidth, trim = 2.5cm 0.5cm 1.8cm 0cm]{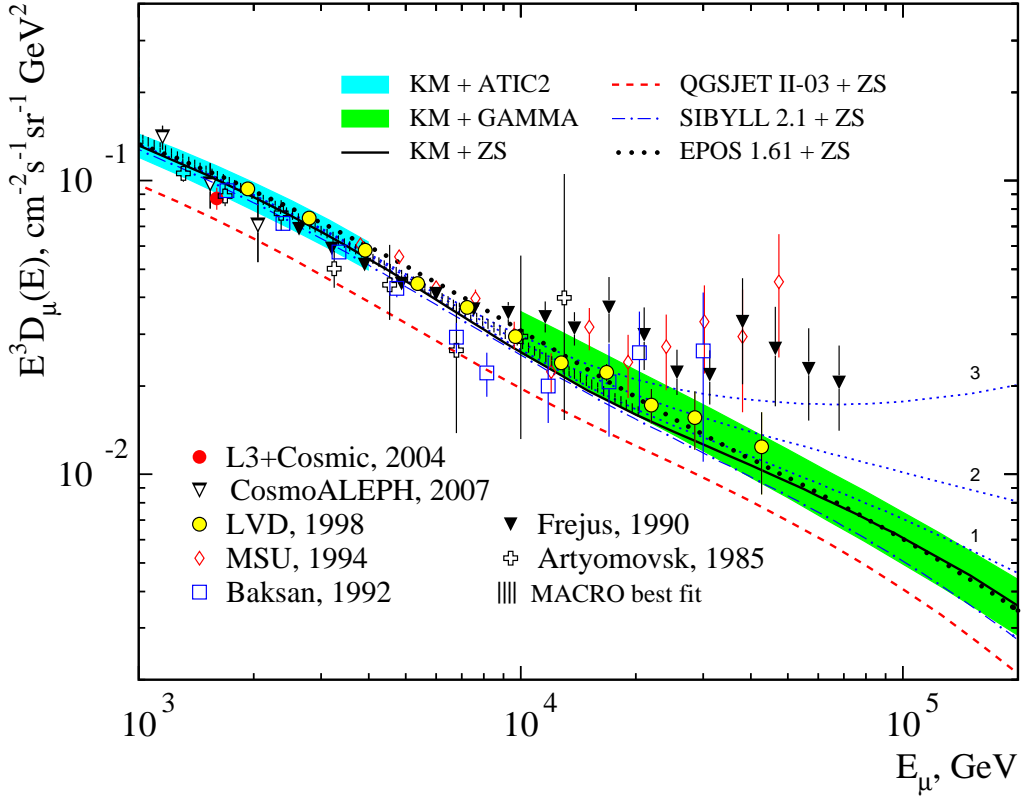}
	\end{center}
	\caption{High-energy plot of the ground level muon spectrum. The solid curves and bands show this work calculations for the KM model with usage of ATIC-2, GAMMA, and ZS primary spectra.
The reults for SIBYLL (dash-dotted line), QGSJET II (dashed), and EPOS (bold dotted) are obtained using the ZS primary spectrum. The dotted lines present sum of the conventional and prompt muon flux (see text for details).}
\label{mu-2}
\end{figure}

 The results of this work calculations of the surface muon flux near the vertical are presented in Figs.~\ref{mu-1} and \ref{mu-2} along with data of the old and recent muon experiments:
the direct measurements of the CAPRICE~\cite{CAPRICE94}, BESS-TeV \cite{BESSTEV}, L3+Cosmic~\cite{L3C}, Cosmo-ALEPH (see Ref.~\cite{Lecoultre}), MASS~\cite{MASS93},  and L3 (taken from~\cite{Bugaev98}) as well as data (converted to the surface) of the underground experiments MSU~\cite{MSU}, MACRO \cite{MACRO}, LVD~\cite{LVD}, Frejus~\cite{Frejus}, Baksan~\cite{Baksan}, and Artyomovsk~\cite{ASD}. 

 The light shaded area in these figures shows the muon flux computed with the KM hadronic model spectrum taking into consideration the uncertainties in the ATIC-2 spectrum data. The dark shaded area on the right in Fig.~\ref{mu-2} shows a high-energy part of the muon flux calculated with the KM model for the GAMMA primary spectrum. The curves represent  muon flux predictions we have made basing on the ATIC-2 primary spectra and using the variety of hadronic models:  solid,  thin, dashed, and bold-dotted lines stand for the calculation with usage of the  KM, SIBYLL, QGSJET II, and EPOS models respectively. Note that at energies beyond the scope of the ATIC-2 experiment, the ZS primary spectrum model was used. The latter fits well the ATIC-2 data and can be considered as a reliable bridge from the TeV to PeV energy range. 

The muon flux predictions obtained with usage of the KM, SIBYLL, and EPOS  models agree well with recent experimental data for energies below $10$ TeV. The range of the muon flux predictions indicates the minimal uncertainty ($\sim 10\%$) due to hadronic models . This uncertainty is comparable with that in the primary cosmic ray spectrum measured in the ATIC-2 experiment. However, in a case of the QGSJET II model the muon flux is $30\%$ lower if compared both with the experimental data and calculations mentioned above.    
The predictions of the AM flux with usage of the GH primary spectrum were obtained for the KM hadronic model in Ref.~\cite{KSS} and for the QGSJET 01, SIBYLL, NEXUS, and QGSJET II models in Ref.~\cite{Lagutin08}. 

The prompt muon component of the flux is shown in Fig.\ref{mu-2} (dotted lines) where   
numbers above the curves indicate the calculations based on the charm production  models: $1$ --     quark-gluon string~\cite{KP, NC89},  $2$ -- recombination quark-parton~\cite{Bugaev98, NC89}, and $3$ -- Volkova et al.~\cite{VFGS}.  These components were added as corrections to the conventional flux (solid line) in order to demonstrate the extent of the  uncertainty  of the calculated muon flux at PeV scale.

\section{Zenith-angle distribution of the muon flux at sea level}

The zenith-angle distribution of the atmospheric muon is of a special interest because it is sensitive to details of the meson production and decays~\cite{ZK60}.
In particular, the $K/\pi$ flux ratio, being dependent both on the hadron model and primary cosmic ray composition, has effect on the muon flux at inclined directions. 
The consistent description of the flux at large zenith angles and at the vertical direction would confirm the adequacy of the hadronic model and the validity of the primary cosmic ray spectrum data. 

\begin{figure}
	\centering
\includegraphics[width=0.85\textwidth, trim = 0.1cm 0.5cm 0.0cm 0.0cm]{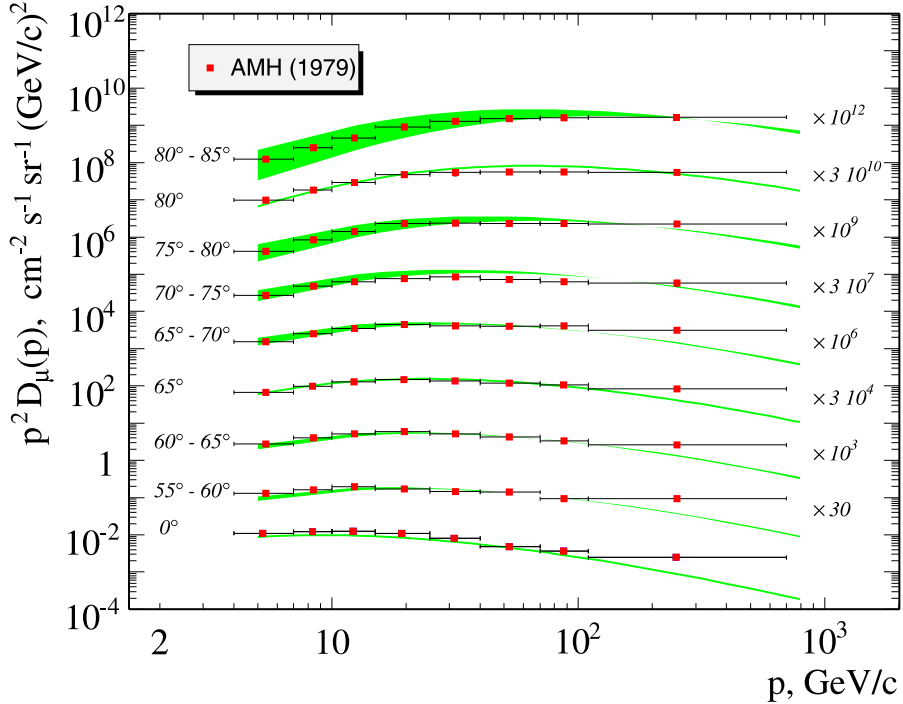}
	\caption{The muon flux calculated for wide range of zenith angles in comparison with the data of AMH~\cite{AMH} magnetic spectrometer. Calculations are performed for the ATIC-2 primary cosmic ray spectra and KM model. The bands show the calculated flux spread due to uncertainties in zenith angles.} 
  		\label{amh}
\end{figure}

\begin{figure}        
	\centering
\vspace{0.51cm}
\includegraphics[width=0.85\textwidth, trim = 0cm 0.5cm 0.3cm 0.0cm]{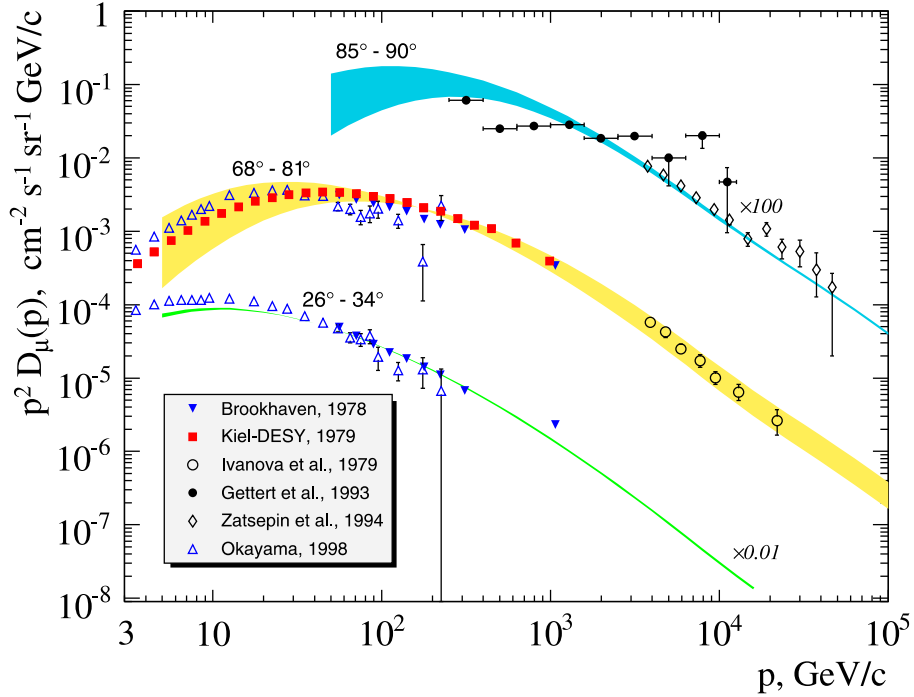}
\caption{Comparison of predicted muon fluxes for inclined directions with the data of  Brookhaven~\cite{BMS}, Kiel-DESY~\cite{Kiel-DESY}, MSU~\cite{MSU, Ivanova},
Okayama~\cite{Okayama} and  Karlsruhe~\cite{Gettert} experiments. Same notations as in Fig.~\ref{amh}.}
\label{Okayama}
\end{figure}

\begin{figure}
	\centering
\includegraphics[width=0.90\textwidth, trim = 0.5cm 0.0cm 0.4cm 0.0cm]{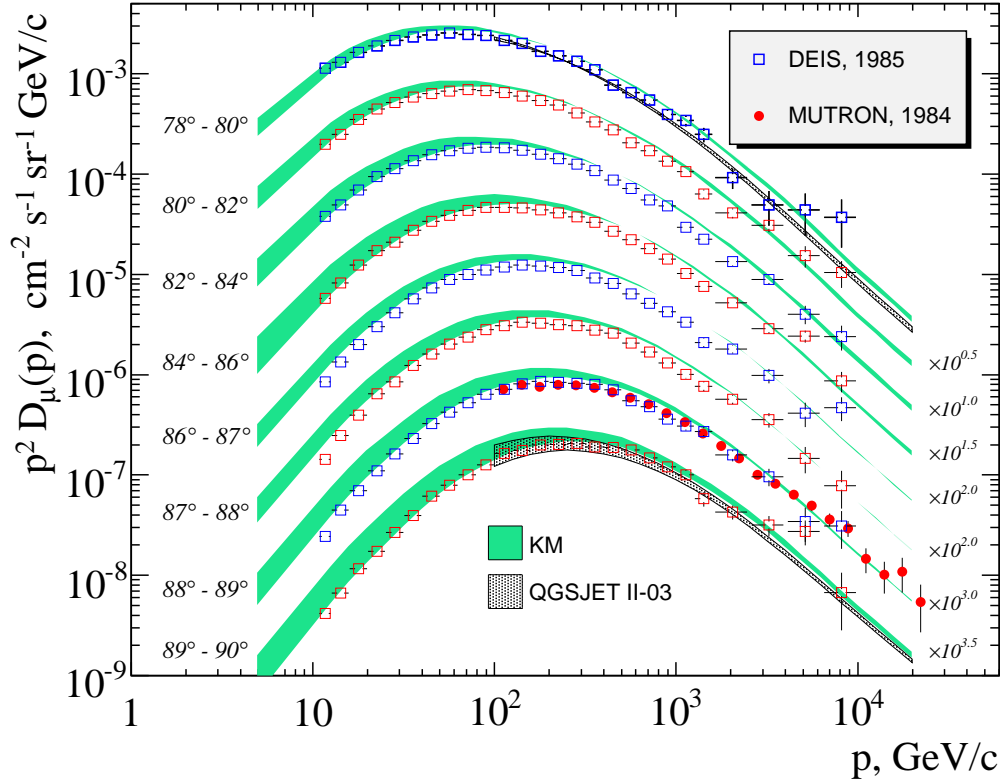}
\caption{Comparison of near-horizontal muon fluxes at sea level with the data 
of MUTRON~\cite{MUTRON}  and  DEIS~\cite{DEIS}. The calculations are made for the ATIC-2 primary spectrum with usage of  KM (shaded areas) and QGSJET II (hatched areas).  
}  
 	\label{deis}
\end{figure}

\begin{figure}
	\centering
\vspace{0.2cm}
\includegraphics[width=0.77\textwidth, trim = 0.5cm 0.3cm 0.4cm 0.5cm]{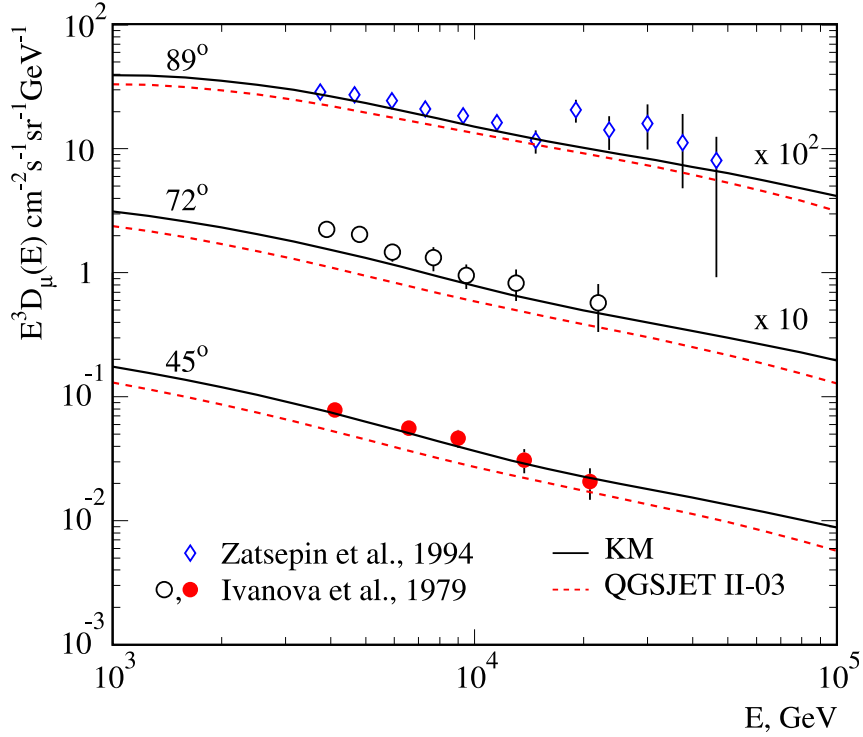}
\caption{Comparison of the oblique muon fluxes at sea level with the data of MSU  experiments~\cite{MSU, Ivanova} (see also Fig.~\ref{Okayama}). The calculations are made for the ATIC-2 primary spectrum using KM model (solid lines) and QGSJET II (dashed).   
}  
 	\label{msu}
\end{figure}

\begin{figure}
	\centering
	\vspace{0.6cm}
\includegraphics[width=0.70\textwidth, trim = 1.5cm 0.3cm 1.0cm 0.0cm]{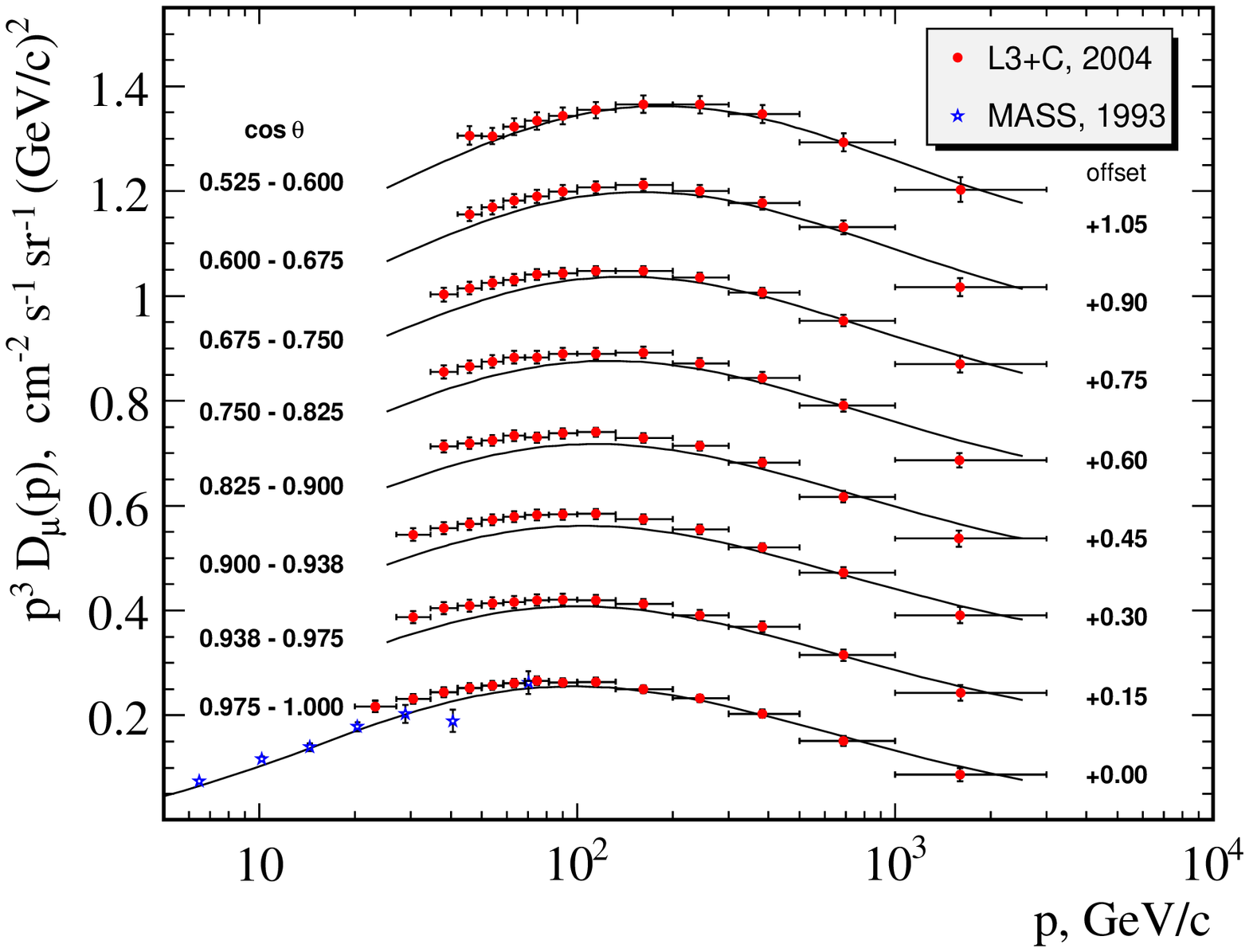}
	\caption{Comparison of the predictions for near-vertical muon fluxes with the data of the L3+Cosmic experiment. The computations are made for the ATIC-2 primary spectrum with usage of  KM.  The MASS data are added to view the low-energy junction.}  
  		\label{l3c}
\end{figure}

A comparison of the measurements in wide zenith angle range with the muon energy spectra  calculated for the ATIC-2 primary cosmic ray spectra and KM model is shown in Figs.~\ref{amh}--\ref{l3c}.
The calculations agree rather well with  old experimental data obtained  with magnetic spectrometers, AMH~\cite{AMH}, BMS~\cite{BMS}, Kiel-DESY~\cite{Kiel-DESY}, MUTRON~\cite{MUTRON}, DEIS~\cite{DEIS} or  X-ray emulsion chambers of MSU~\cite{MSU, Ivanova}, as well as with comparatively recent results obtained by Okayama telescope~\cite{Okayama}, those of the Karlsruhe experiment~\cite{Gettert} and the L3+Cosmic ~\cite{L3C}. 



Figure~\ref{deis} shows the reliable measurements with the magnetic spectrometers DEIS and MUTRON for zenith angles ranging from $78^\circ$ to $90^\circ$ and our predictions based on the KM model(shaded areas) and QGSJET II (hatched area). For clearness, the QGSJET II predictions are shown only for the two angle ranges $78^\circ$-$80^\circ$ and $89^\circ$-$90^\circ$. Unlike the vertical direction,  the muon data for large zenith angles are described by QGSJET II  rather well.  A difference between the QGSJET II and KM models  reduces now to $15$--$20\%$.
Muon fluxes at angles of $45^\circ$, $72^\circ$, and $89^\circ$  were measured by the MSU group~\cite{MSU, Ivanova} using with X-ray emulsion chambers (see Fig.~\ref{deis},~\ref{msu}).
The calculations done using the ZS primary cosmic-ray spectrum and KM model agree well  with the MSU data including horizontal directions.  It is evident from  Fig.~\ref{msu} that the difference between the QGSJET II and KM models decreases approximately from $30\%$ to $15\%$ as zenith angle increases. 
To compare the predictions with the recent L3+Cosmic experiment data for the zenith-angle range $0^\circ \leq\theta\leq 58^\circ$ (Fig.~\ref{l3c}), we apply the atmospheric density profile measured near the site of L3+Cosmic experiment. Apparently our calculation agrees with the L3+Cosmic data at $p_{\mu}\gtrsim 50$ GeV/c. Solid curves correspond to the average zenith angles for a given angle bin. Also for comparison, the low-energy data of the MASS experiment~\cite{MASS93} are added to the figure.

\begin{figure}
	\centering
\includegraphics[width=0.95\textwidth, trim = 2.0cm 0.5cm 1.5cm 0.0cm]{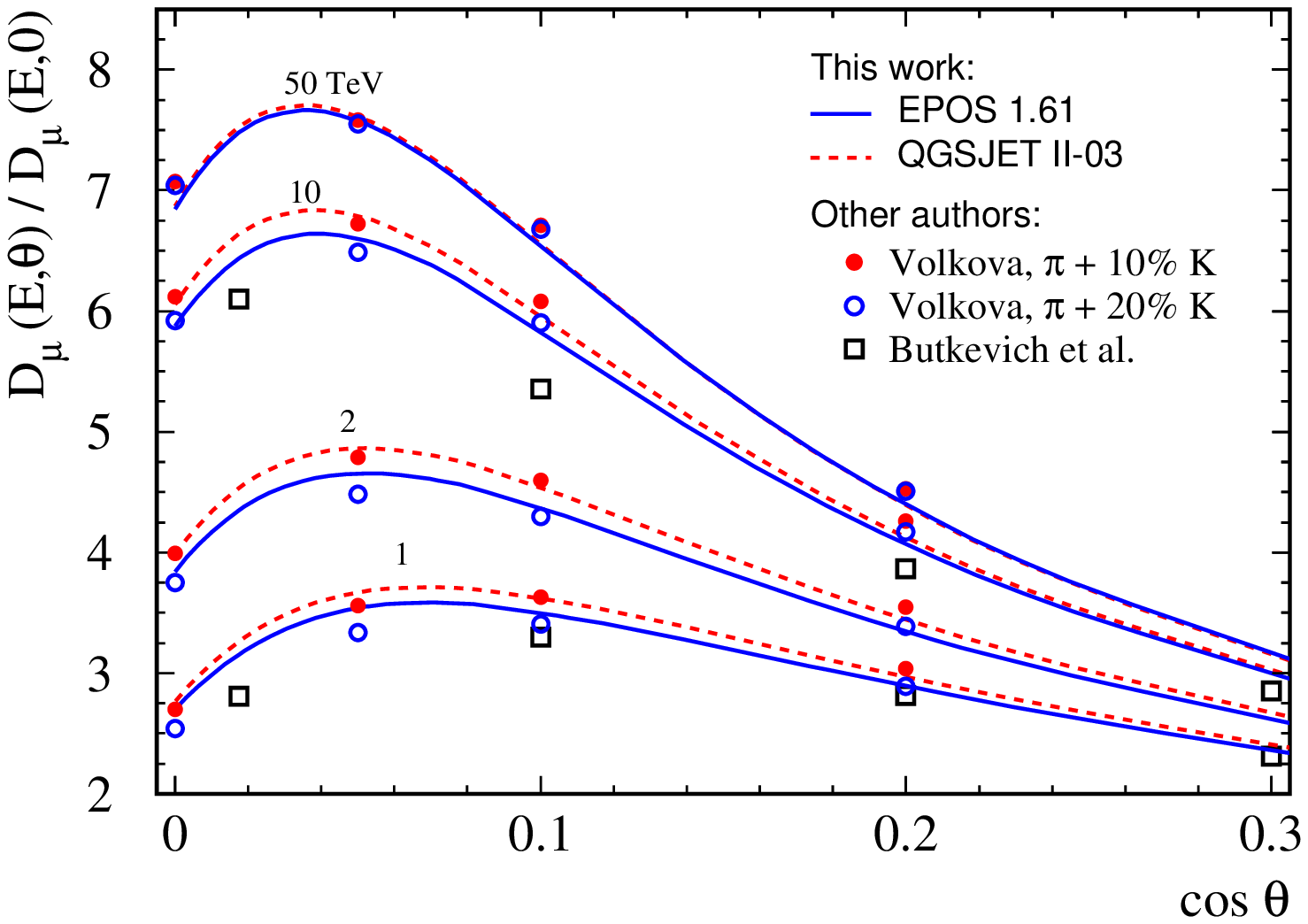}
\caption{Zenith-angle enhancement of the muon flux, $D_{\mu}(E, \theta)/D_{\mu}(E, 0^\circ)$, calculated with the EPOS (solid lines) and  QSJET II (dashed), both for the ATIC-2 primary spectrum. Symbols present the calculations of ~\cite{Dedenko89}(squares) and \cite{Amineva} (circles). Numbers above curves are the muon energies in TeV.}  
        \label{cosmu_en}
\end{figure}

The effect of $K/\pi$ ratio on the zenith-angle distribution of high-energy muons is rather transparent:  
each new source of the muon flux leads to decrease of the muon flux ratio,  $D_{\mu}(\theta)/D_{\mu}(0)$~\cite{ZK60, Amineva}. The more intensive is source (i.e. the larger $K/\pi$ ratio in this case), the lesser is the angle enhancement of the muon flux.
However we can observe the influence of  $K/\pi$ ratio (that is the proper feature of the hadron model) only in the restricted energy region, because the ratio $D_{\mu}(\theta)/D_{\mu}(0)$ weakly depends on the $K/\pi$ ratio for
 $E \gg \epsilon_K(\theta)$, where $\epsilon_K(\theta)$ is the critical energy for kaons (e.g., $\epsilon_{K^\pm}(0)\simeq \,850 {\rm GeV}$). The reason is that the meson decay probability decreasing with the energy,
 $w_d(E, \theta)\propto \epsilon_{\pi, K}(\theta)/E$, leads to $D_{\mu}(\theta)/D_{\mu}(0) \propto \epsilon_{\pi, K}(\theta)/ \epsilon_{\pi, K}(0)$,  or $D_{\mu}(\theta)/D_{\mu}(0)\simeq 7.5$ near the maximum of the muon zenith-angle distribution 
 (see Fig.~\ref{cosmu_en}).

Fig.~\ref{cosmu_en} shows the muon flux ratio $D_{\mu}(\theta)/D_{\mu}(0)$ as a function of $\cos\theta$, that is a coefficient of zenith-angle enhancement of the muon flux, calculated with the EPOS model (solid lines) and  QSJET II one (dashed) in the range $1$-$50$ TeV.
A choice of these hadronic models is motivated by an observation that the EPOS and QGSJET II lead to the outermost predictions of the hadron fluxes and  $K/\pi$ ratio (see Fig.~\ref{PAMIR} and~\ref{kpiratio}). This gives a possibility to examine indirectly the effect of the kaon source on the muon angular flux: the EPOS prediction of the  $K/\pi$ ratio ($\sim 0.2$ at $E = 10$ TeV) results in lower  $D_{\mu}(\theta)/D_{\mu}(0)$ ratio in comparison with that of QGSJET II model for which the $K/\pi$ ratio is close to $0.1$ at the same energy (compare also with predictions by Volkova~\cite{Amineva} for $K/\pi=0.1$, closed circles in Fig.~\ref{cosmu_en}).


\begin{figure}
	\centering
\includegraphics[width=0.95\textwidth, trim = 2.3cm 0.2cm 1.9cm 0.0cm]{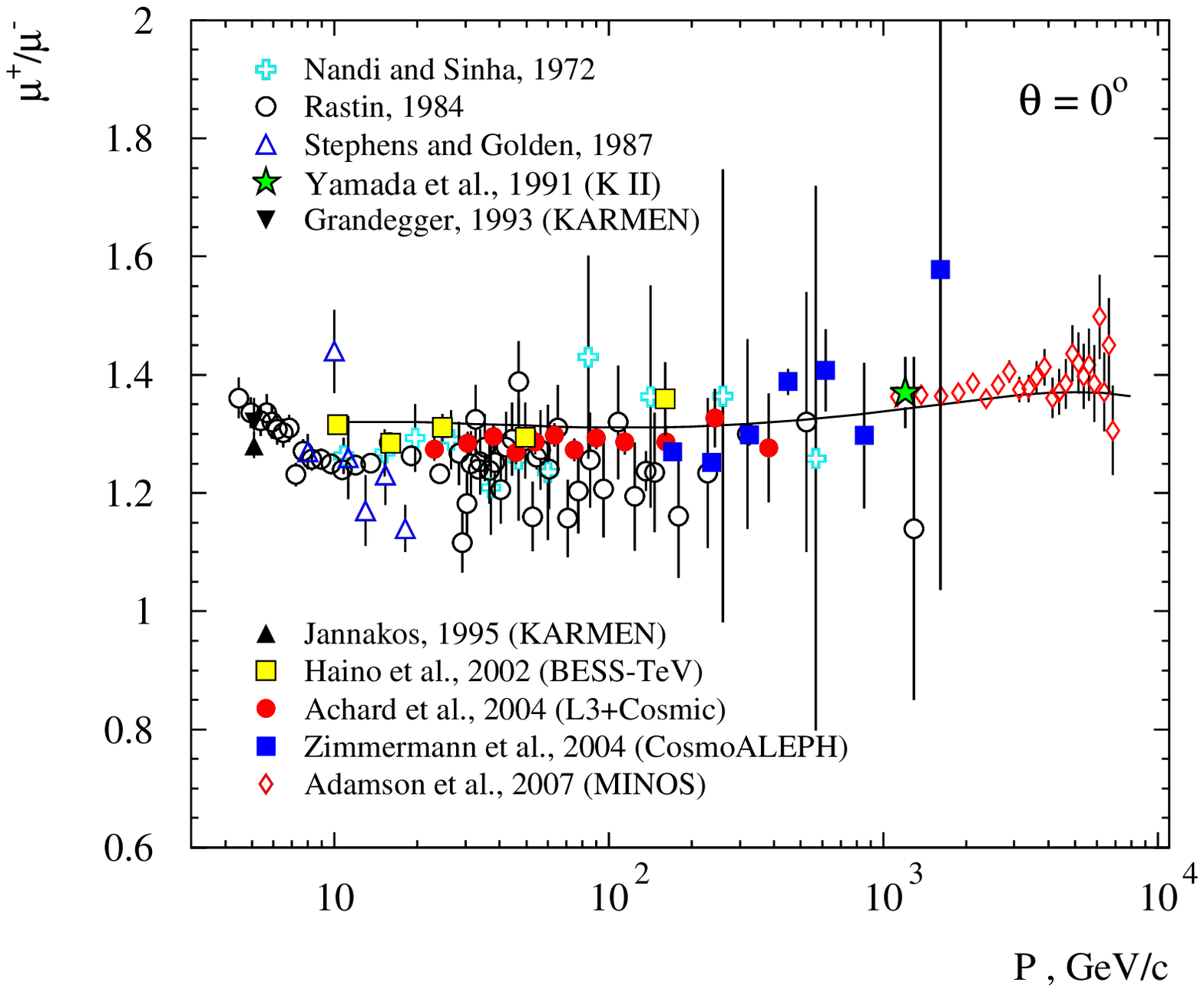}
\caption{Muon charge ratio at sea level calculated with KM model for the ATIC-2 primary  spectra. The data of experiments are taken from Refs.~\cite{BESSTEV, L3C, Nandi, Rastin, Stephens, Jannakos, Zimmermann, MINOS, kamioka}.
}  
  		\label{charge_ratio} \vspace{6 pt}
\end{figure}

A comparison of the calculated $\mu^+/\mu^-$  ratio with measurements in a wide energy range makes a possibility to study hadron-nucleus interactions indirectly. Present computation (solid line) of the muon charge ratio $\mu^+/\mu^-$  at sea level  is plotted in Fig.~\ref{charge_ratio} along with experimental data from Refs.~\cite{Nandi, Rastin, Stephens, Jannakos, Zimmermann, MINOS, kamioka}.  This computation has been performed with the KM hadronic model and the ATIC-2 primary spectrum. For muon energies below the charged pion critical energy ($\sim 115$ GeV), the pions are the mostly dominant source of muons. For these energies, the calculated ratio, $\sim 1.3$,  agrees with the recent data of BESS-TeV, CosmoALEPH, and  L3+Cosmic experiments. As energy increases the kaon source of muons intensifies leading to a maximum of the muon ratio,  $\sim 1.4$,  at energy close to $10$ TeV. This value agrees with the most recent results obtained with the MINOS far detector~\cite{MINOS}.

\section{Summary}

In this work, we have calculated the atmospheric high-energy fluxes of  nucleons, mesons and muons implementing the hadronic models QGSJET~\cite{QGSJET01, QGSJET2}, SIBYLL~\cite{SIBYLL}, NEXUS~\cite{NEXUS}, EPOS~\cite{EPOS}, and  the Kimel and Mokhov one ~\cite{KMN}.  These calculations basing on the method~\cite{NS, KSS} have been performed using directly data for primary spectra and composition of the ATIC-2 experiment~\cite{ATIC2},  GAMMA one~\cite{GAMMA} and Zatsepin and Sokolskaya model~\cite{ZS3CM}. Alternatively  we used well known  parameterization by Gaisser and Honda~\cite{GH}. The method appears rather efficient tool to study the transport of cosmic ray particles through the atmosphere in a generalized case of the non-power law primary spectrum and non-scaling behavior of hadronic cross sections. Various hadronic models can be rather easily embedded into the developed code.

The present calculations demonstrate the consistency of the new primary comic ray spectra measurements with recent experimental data on atmospheric hadron and muon fluxes at different levels of atmosphere for a wide range of energies and zenith angles. The comparison of the calculated and measured fluxes gives the possibility to evaluate the uncertainties originated from the errors of the primary spectrum data. 
The high accuracy of the ATIC-2 data resulted in the muon flux calculation uncertainty, comparable with rather high precision of the last decade muon flux measurements.
Also this allowed us to study the effect of the non-power law structure of the ATIC-2 primary spectrum on secondary fluxes. The effect on the muon spectrum shape of deviations from a power law in Zatsepin and Sokolskaya primary spectrum appears at energy above $10$ TeV. 

The vertical muon flux predictions, obtained with usage of the  ATIC-2 primary spectrum and KM, SIBYLL, EPOS hadronic models, are in accordance with recent experimental data in the wide energy range. These models give similar muon fluxes differing by $\sim 10\%$, that is comparable to the uncertainty in the ATIC-2 data. The muon spectrum predicted with the QGSJET II model is $\sim 30\%$ lower in the energy range $10^2-10^5$ GeV if compared both with the experimental data and calculations with usage of the rest models.  For near horizontal directions the uncertainties in the muon flux calculations are found to be less in comparison with that for the vertical. In particular, the  discrepancy between the QGSJET II and KM models is reduced to something like $15\%$.  Thus we may state: the uncertainties due to hadronic cross sections are about $30\%$, while those from variations in the primary spectrum are at a level of $10\%$.

And at last, the prompt muon component originated from the decay of the charmed hadrons is still not be reliably extracted from the experiments, and it seems to be the source of hardly estimated uncertainty that superimposes on the uncertainty factor due to 
poor knowledge of the primary cosmic ray flux around the knee.

 \section{ACKNOWLEDGMENTS}
 
We are grateful to V.~A.~Naumov for enlightening discussions on this study. We are much obliged to T.~Pierog for clarifying comments that facilitated to implement the codes of the modern hadronic interaction models.
This work was partly supported by Federal Programme ``Russian Scientific Schools", grant NSh-1027.2008.2. 


\begin{thebibliography}{00}

\bibitem{QGSJET01}
N.~N.~Kalmykov, S.~S.~Ostapchenko, A.~I.~Pavlov, Nucl. Phys. B (Proc. Suppl.) 52 (1997) 17. 

\bibitem{QGSJET2}
S.~S.~Ostapchenko, Nucl. Phys. B (Proc. Suppl.) 151  (2006) 143;\\
S.~Ostapchenko, Phys. Rev. D 74 (2006) 014026.

\bibitem{ostap08}
S.~Ostapchenko, Nucl. Phys. B (Proc. Suppl.) 175-176 (2008) 73.

\bibitem{SIBYLL}
R.~S.~Fletcher, T.~K.~Gaisser, P.~Lipari, T.~Stanev, Phys.Rev. D 50 (1994) 5710; \\
R.~Engel, T.~K.~Gaisser, P.~Lipari, T.~Stanev, in: Proc. 26th ICRC, Salt Lake City, 1999,  vol. 1, p. 415.

\bibitem{NEXUS}
H.~J.~Drescher et al., Phys. Rep. 350  (2001) 93; \\
T.~Pierog et al., Nucl. Phys. A 715  (2003) 895;\\
K.~Werner et al.,  J. Phys. G: Nucl. Part. Phys. 30 (2003) S211.

\bibitem{EPOS}
K.~Werner, T.~Pierog, AIP Conf. Proc. 928 (2007) 111. Available from: $<$ astro-ph/0707.3330$>$; \\
T.~Pierog, K.~Werner, in: Proc. 30th ICRC, Merida, HE.1.6, No.905, 2007;\\
K.~Werner, Nucl. Phys. B (Proc. Suppl.) 175-176 (2008) 81.

\bibitem{KMN}
A.~N.~Kalinovsky, N.~V.~Mokhov, Yu.~P.~Nikitin, Passage of high-energy particles through matter, AIP, New York,  1989.

\bibitem{ATIC2}
A.~D.~Panov et al., Bull. Russ. Acad. Sci. Phys. 71 (2007) 494.  Available from:$<$astro-ph/0612377$>$.

\bibitem{GAMMA}
A.~P.~Garyaka et al., Astropart. Phys. 28 (2007) 169; \\ 
S.~V.~Ter-Antonyan et al., in: Proc. 29th ICRC, Pune, vol. 6, 2005, p. 101. 
Available from:$<$astro-ph/0506588$>$.

\bibitem{ZS3CM}
V.I.~Zatsepin, N.V.~Sokolskaya, A\&A 458 (2006) 1; Astron. Lett. 33 (2007) 25.

\bibitem{GH}
T.~K.~Gaisser, M.~Honda, Annu. Rev. Nucl. Part. Sci. 52 (2002) 153. \\ 
T.~K.~Gaisser, M.~Honda, P.~Lipari, T.~Stanev, in: Proc. 27th ICRC, Hamburg, 2001, vol. 1, p. 1643. 

\bibitem{CAPRICE94} 
J.~Kremer \emph{et al.}, Phys. Rev. Lett. 83 (1999) 4241. 


\bibitem{BESSTEV}
S.~Haino et al., Phys. Lett. B 594 (2004) 35.

\bibitem{L3C}
P.~Achard et al., Phys. Lett. B 598  (2004) 15.

\bibitem{Lecoultre}
P.~Le~Coultre, in: Proc. 29th ICRC, Pune, 2005, vol. 10, p. 137.

\bibitem{Naumov99}
V.~A.~Naumov, L.~Perrone, Astropart. Phys. 10 (1999) 239.

\bibitem{NS}
V.~A.~Naumov,  T.~S.~Sinegovskaya, Phys. Atom. Nucl. 63 (2000) 1927;
in: Proc. 27 ICRC, Hamburg, 2001, Vol. 1, p. 4173. Available from: $<$hep-ph/0106015$>$.

\bibitem{KSS}
A.~A.~Kochanov, T.~S.~Sinegovskaya, S.~I.~Sinegovsky, Phys. Atom. Nucl. 70 (2007) 1913.

\bibitem{KPSS}
A.~A.~Kochanov, A.D.Panov, T.~S.~Sinegovskaya, S.~I.~Sinegovsky, in: Proc. of 30 ICRC,  Merida, 2007, HE.2.4 521. Available from: $<$astro-ph/0706.4389$>$.

\bibitem{ZK60} 
G.~T.~Zatsepin, V.~A.~Kuzmin, Zh. Eksp. Teor. Fiz. 39 (1960) 1677.

\bibitem{VZK79}
L.~V.~Volkova, G.~T.~Zatsepin, L.~A.~Kuzmichev, Yad. Fiz. 29 (1979) 1252.

\bibitem{Dedenko89}
A.~V.~Butkevich, L.~G.~Dedenko, I.~M.~Zheleznykh, Sov. J. Nucl. Phys. 50 (1989) 90.

\bibitem{Lipari}
P.~Lipari, Astropart. Phys. 1 (1993) 195.

\bibitem{Bugaev98}
E.~V.~Bugaev et al., Phys. Rev. D 58  (1998) 054001. Available from: $<$hep-ph/9803488$>$.

\bibitem{Naumov2001}
G.~Fiorentini, V.~A.~Naumov, F.~L.~Villante, Phys. Lett. B 510 (2001) 173. 

\bibitem{VZ01}
L.~V.~Volkova, G.~T.~Zatsepin, Phys. Atom. Nucl. 64 (2001) 266.

\bibitem{Ryazhskaya96}
O.~G.~Ryazhskaya, Nuovo Cim. C 19 (1996) 655.

\bibitem{Costa}
C.~G.~S.~Costa,  Astropart. Phys. 16 (2001) 193.

\bibitem{Naumov2004}
V.~A.~Naumov, in: Proc. 2nd Workshop on method. aspects of underwater/ice neutrino telescopes, Hamburg, 2001. Ed. by R. Wischnewski (DESY, Hamburg, 2002), p. 31. 
Available from: $<$hep-ph/0201310$>$.

\bibitem{Hebbeker02}
T.~Hebbeker and C.~Timmermans, Astropart. Phys. 18 (2002) 107. 

\bibitem{BESS98}
T.~Sanuki et al.,  Astrophys. J. 545 (2000) 1135.

\bibitem{AMS}
J.~Alcaraz et al.,  Phys. Lett. B 490 (2000) 27; \\ 
J.~Alcaraz et al.,  ibid. 494 (2000) 193.

\bibitem{IMAX}
W.~Menn et al.,  Astrophys. J. 533  (2000) 281.


\bibitem{CAP98} 
M.~Boezio et al.  Astropart. Phys. 19 (2003) 583; \\
M.~Boezio et al.  Astrophys. J. 518  (1999) 457.

\bibitem{MASS}
R.~Bellotti et al., Phys. Rev. D 60 (1999) 052002.

\bibitem{RICH}
J.~Buckley et al.,  Astrophys. J. 429 (1994) 736; \\
T.~T.~von~Rosenvinge, in:  Proc. 25th ICRC, Durban, 1997, vol. 8, p. 237. 

 \bibitem{MUBEE}
V.~I.~Zatsepin et al., Phys. Atom. Nucl. 57 (1994) 645.

\bibitem{RUNJOB}
A.~V.~Apanasenko et al., Astropart. Phys. 16 (2001) 13.

\bibitem{JACEE}
K.~Asakimori et al., Astrophys. J. 502 (1998) 278.

\bibitem{SOKOL}
I.~P.~Ivanenko et al., in: Proc. 23rd ICRC, Calgary, 1993,  vol. 2, p. 17.

\bibitem{KASSH}
T.~Antoni et al., Astrophys. J. 612  (2004) 914.

\bibitem{KAS_elemnt}
T.~Antoni et al., Astropart. Phys.  24  (2005) 1.

\bibitem{HEGRA}
F.~Aharonian et al., astro-ph/9901160.

\bibitem{TIBETHD}
M.~Amenomori et al., Phys. Rev. D 62 (2000) 112002.

\bibitem{ICHIMURA}
M.~Ichimura et al., Phys. Rev. D 48 (1993) 1949.

\bibitem{Berezhko}
E.~G.~Berezhko, H.~J.~V$\ddot{\text o}$lk,  
Astrophys. J. Lett. 661 (2007) L175.
Available from:  $<$astro-ph/0704.1715$>$ 

\bibitem{CORSIKA}
D.~Heck et al., FZKA Report 6019 (Forschungszentrum Karlsruhe GmbH, Karlsruhe, 1998).

\bibitem{aires}
S.~J.~Sciutto, AIRES - a system for air shower simulations. User's reference manual (LaPlata, Argentina, 2002) pp. 1-250. Available from: $<$http://www.fisica.unlp.edu.ar/auger/aires/$>$; \\
S.~J.~Sciutto, in Proc. 26th ICRC, Salt Lake City, Utah, 1999, vol. 1, 411. Available 
from: $<$astro-ph/9905185$>$.
 
\bibitem{CONEX}
T.~Pierog et al., Nucl. Phys. B (Proc. Suppl.) 151 (2006) 159; \\ 
T.~Bergmann et al., Astropart. Phys. 26 (2007) 420. Available from: $<$astro-ph/0606564$>$.

\bibitem{Derome}
 L.~Derome,  Phys. Rev.  D 74 (2006) 105002.

\bibitem{Mielke}
H.~H.~Mielke et al., J. Phys. G 20 (1994) 637.

\bibitem{Baltrusaitis}
R.~M.~Baltrusaitis et al., Phys. Rev. Lett. 52 (1984) 1380.

\bibitem{Honda}
M.~Honda et al., Phys. Rev. Lett. 70 (1993) 525.

\bibitem{Yakutsk}
S.~P.~Knurenko et al., in:  Proc. 26th ICRC, Salt Lake City, Utah, vol. 1, 1999, p. 372.

\bibitem{HiRes}
K.~Belov et al., in: 30th ICRC, Merida, HE.3.1, No.1216, 2007.

\bibitem{EASTOP1}
M.~Aglietta et al., in: 30th ICRC, Merida, HE.1.1.A, No.589, 2007.

\bibitem{ARGO}
I.~De Mitri et al., in: 30th ICRC, Merida, HE.3.1, No.950, 2007.

\bibitem{EGLS92}
J.~Engel, T.~K.~Gaisser, P.~Lipari, T.~Stanev, Phys.Rev. D 46 (1992) 5013. 

\bibitem{Lipari08}
P.~Lipari, Nucl. Phys. B (Proc. Suppl.) 175-176 (2008) 96.

\bibitem{NC98}V.~A.~Naumov, T.~S.~Sinegovskaya, S.~I.~Sinegovsky, Nuovo Cim. A  111 (1998) 129.
Available from: $<$hep-ph/9802410v3$>$.

\bibitem{NCC98}
V.~A.~Naumov, T.~S.~Sinegovskaya, S.~I.~Sinegovsky, in: Proc. Baikal School for Young Researchers on the Astrophysics and Microworld Physics, Irkutsk, Russia, 1998, p. 211 (in Russian). Available from: $<$http://bsfp.iszf.irk.ru/bsfp1998/Reports/kl3.ps$>$.

\bibitem{VNS}
A.~N.~Vall, V.~A.~Naumov, S.~I.~Sinegovsky, Sov. J. Nucl. Phys. 44 (1986) 806.

\bibitem{Malhotra}
P.~K.~Malhotra et al., Nature 209 (1966) 567.

\bibitem{Apanasenko}
A.~V.~Apanasenko, M.~N.~Sherbakova, Izv. Akad. Nauk Ser. Fiz. 32 (1968) 372.

\bibitem{Aguirre}
C.~Aguirre et al. , Can. J. Phys. 46 (1968) S660.

\bibitem{KASCADE}
H.~Kornmayer et al., J. Phys. G 21 (1995) 439. 

\bibitem{Kuzmichev}
L.~A.~Kuzmichev et al., in: Proc. 17th ICRC, Paris, vol. 2, 1981, p. 103.

\bibitem{Takahashi}
Y.~Takahashi, in: Proc. 16th ICRC, Kyoto, vol. 7, 1979, p. 115.

\bibitem{Malinowski}
J.~Malinowski et al., in: Proc. 29th ICRC, Pune, vol. 6, 2005, p. 61.

\bibitem{ThienShan}
S.~D.~Adamov et al., in: Proc. of  18th ICRC, Bangalore, vol. 5, 1983, p. 275.

\bibitem{Erofeeva}
I.~N.~Erofeeva, in: Mater. Vsesoyuz. konf. po kosmicheskim lucham, Tashkent, 1968, (FIAN, Moscow, 1969) part 1, issue 1, p. 92.

\bibitem{Babayan}
Kh.~P.~Babayan et al., Izv. Akad. Nauk Ser. Fiz. 31 (1967) 1425.

\bibitem{EASTOP}
M.~Aglietta et al., Astropart. Phys. 19 (2003) 329. 

\bibitem{Shmeleva}
P.~A.~Shmeleva et al., in: Proc. 18th ICRC, Bangalore, vol. 5, 1983, p. 271.

\bibitem{Ashton}
F.~Ashton and A.~J.~Saleh, Nature 256 (1975) 387.


\bibitem{MASS93}
M.~P.~de Pascale et al., J. Geophys. Res. 98 (1993) 3501.

\bibitem{MSU}
G.~T.~Zatsepin et al., Izv. Ross. Akad. Nauk Ser. Fiz. 58 (1994) 119.

\bibitem{MACRO} 
M.~Ambrosio et al., Phys.Rev. D 52 (1995) 3793.

\bibitem{LVD}
M.~Aglietta et al., Phys. Rev. D 58 (1998) 092005.

\bibitem{Frejus}
W.~Rhode, Nucl. Phys. B (Proc. Suppl.) 35 (1994) 250.

\bibitem{Baksan}
V.~N.~Bakatanov et al., Yad. Fiz. 55 (1992) 2107.

\bibitem{ASD}
R.~I.~Enikeev et al., Yad. Fiz. 47  (1988) 1044.

\bibitem{Lagutin08}
A.~V. Yushkov, A.~A.~Lagutin, Nucl.Phys. B (Proc. Suppl.) 175-176 (2008) 170.

\bibitem{KP}
A.~B.~Kaidalov, O.~I.~Piskunova, Sov. J. Nucl. Phys. 41 (1985) 816; Sov. J. Nucl. Phys. 43 (1986) 994; Z. Phys. C30 (1986) 145; \\ 
O.~I.~Piskunova, Sov. J. Nucl. Phys. 56 (1993) 1094. 

\bibitem{NC89}
E.~V.~Bugaev et al., Il Nuovo Cimento  C 12 (1989) 41. 

\bibitem{VFGS}
L.~V.~Volkova et al, Il Nuovo Cimento C 10 (1987) 465.

\bibitem{AMH}
P.~J.~Green et. al., Phys. Rev. D 20 (1979) 1598.

\bibitem{BMS}
R.~G.~Kellogg, H.~Kasha, Phys. Rev. D 17 (1978) 98.

\bibitem{Kiel-DESY}
H.~Jokisch et al., Phys. Rev. D 19 (1979) 1368.

\bibitem{MUTRON}
S.~Matsuno et al., Phys. Rev. D 29 (1984) 1.

\bibitem{DEIS}
O.~C.~Allkofer et al., Nucl. Phys. B 259 (1985) 1.

\bibitem{Ivanova}
M.~A.~Ivanova et al., in: Proc. 16th ICRC, Kyoto, vol. 10, 1979, p. 35.

\bibitem{Okayama}
S.~Tsuji et al.,  J. Phys. G: Nucl. Part. Phys. 24 (1998) 1805.

\bibitem{Gettert}
M.~Gettert et al., in: Proc. 23rd ICRC, Calgary, vol. 4, 1993, p. 394.

\bibitem{Amineva}
L.~V.~Volkova, FIAN Report 72 (Lebedev Physical Institute, AN USSR), Moscow, 1969; \\
T.~P.~Amineva et al., Issledovanie muonov sverkhvysokikh energij: Metod rentgenoemulsionnykh kamer. Ed. by G.T. Zatsepin 
Nauka, Moscow, 1975.

\bibitem{Nandi}
B.~C.~Nandi and M.~S.~Sinha, J. Phys. A 5  (1972) 1384.

\bibitem{Rastin}
B.~C.~Rastin, J. Phys. G10  (1984) 1629.

\bibitem{Stephens}
S.~A.~Stephens, R.~L.~Golden, in: Proc. 20th ICRC, Moscow, vol. 6, 1987, p. 173.

\bibitem{Jannakos}
T.~E.~Jannakos, FZKA Report 5520 (Forschungszentrum Karlsruhe, 1995).

\bibitem{Zimmermann}
D.~Zimmermann et al., Nucl. Instrum. Methods A 525 (2004) 141.

\bibitem{MINOS}
P.~Adamson et al., Phys. Rev. D 76 (2007) 052003.

\bibitem{kamioka}
M.~Yamada et al., Phys. Rev. D 44 (1991) 617.

\end{thebibliography}
\end{document}